\documentclass[a4paper,twocolumn,9pt,aps,prb,accepted=2023-05-15]{quantumarticle}
\pdfoutput=1
\usepackage[utf8]{inputenc}
\usepackage[english]{babel}
\usepackage[T1]{fontenc}
\usepackage{tikz}
\usepackage{lipsum}

\usepackage[numbers,sort&compress]{natbib}

\usepackage{amsmath,amssymb}
\usepackage{graphicx}
\usepackage{dcolumn}
\usepackage{enumitem}
\usepackage[normalem]{ulem}
\usepackage{float}
\usepackage{bm,dsfont}
\usepackage{multirow}
\usepackage{simpler-wick}
\usepackage{cuted}
\usepackage{color}
\usepackage{changes}
\usepackage{hyperref}
\hypersetup{
	colorlinks = true,
	linkcolor = [rgb]{0.70,0.13,0.13},
	citecolor = [rgb]{0.13,0.55,0.13},
	urlcolor = [rgb]{0.25, 0.41, 0.88}}
\newcommand{\bra}[1]{{\langle #1|}}
\newcommand{\ket}[1]{{|#1 \rangle}}
\newcommand{\braket}[2]{{\langle #1|#2\rangle}}

\newcommand{\id}{\mathds{1}}

\newcommand{\scE}{\mathcal{E}}

\newcommand{\Tr}{\operatorname{Tr}}

\newcommand{\eq}[1]{\begin{equation}#1\end{equation}}

\newcommand{\eqnref}[1]{Eq.\,\eqref{#1}}
\newcommand{\figref}[1]{Fig.\,\ref{#1}}
\newcommand{\tabref}[1]{Tab.\,\ref{#1}}
\newcommand{\secref}[1]{Sec.\,\ref{#1}}\newcommand{\appref}[1]{Appendix\,\ref{#1}}
\newcommand{\refcite}[1]{Ref.\,\cite{#1}}

\usepackage[mathlines]{lineno}

\begin{document}
\title{Scalable and Flexible Classical Shadow Tomography with Tensor Networks}
\author{A. A. Akhtar}
\affiliation{Department of Physics, University of California San Diego, La Jolla, CA 92093, USA}
\email{a1akhtar@ucsd.edu}
\orcid{0000-0001-5339-0194}
\author{Hong-Ye Hu}
\affiliation{Department of Physics, University of California San Diego, La Jolla, CA 92093, USA}
\affiliation{Department of Physics, Harvard University, 17 Oxford Street, Cambridge, MA 02138, USA}
\email{hongyehu@g.harvard.edu}
\orcid{0000-0001-5841-831X}
\author{Yi-Zhuang You}
\affiliation{Department of Physics, University of California San Diego, La Jolla, CA 92093, USA}
\email{yzyou@physics.ucsd.edu}
\maketitle

\begin{abstract}
Classical shadow tomography is a powerful randomized measurement protocol for predicting many properties of a quantum state with few measurements. Two classical shadow protocols have been extensively studied in the literature: the single-qubit (local) Pauli measurement, which is well suited for predicting local operators but inefficient for large operators; and the global Clifford measurement, which is efficient for low-rank operators but infeasible on near-term quantum devices due to the extensive gate overhead. In this work, we demonstrate a scalable classical shadow tomography approach for generic randomized measurements implemented with finite-depth local Clifford random unitary circuits, which interpolates between the limits of Pauli and Clifford measurements. The method combines the recently proposed locally-scrambled classical shadow tomography framework with tensor network techniques to achieve scalability for computing the classical shadow reconstruction map and evaluating various physical properties. The method enables classical shadow tomography to be performed on shallow quantum circuits with superior sample efficiency and minimal gate overhead and is friendly to noisy intermediate-scale quantum (NISQ) devices. We show that the shallow-circuit measurement protocol provides immediate, exponential advantages over the Pauli measurement protocol for predicting quasi-local operators. It also enables a more efficient fidelity estimation compared to the Pauli measurement.

\end{abstract}



\section{Introduction}

Quantum technology is advancing rapidly. A central task is to characterize and exploit the features of many-qubit quantum states created in the lab \cite{Jamesquant-ph/0103121, Cavesquant-ph/0104088, DArianoquant-ph/0610058, Guta1809.11162}. To fully determine the density matrix $\rho$ of a quantum system of $n$ qubits, exponential ($\sim4^n$) amount of repeated measurements and classical processing is needed \cite{Flammia1205.2300, Haah1508.01797, ODonnell1508.01907}. Therefore full quantum state tomography is not scalable to large systems. However, for many purposes (e.g.~estimating physical observables on the quantum state), a far less complete description is adequate \cite{Aaronson1711.01053, Aaronson1904.08747}, and the amount of measurement and classical processing can be drastically reduced. Much of the recent progress has been made by exploiting the randomized measurement strategy \cite{Ohliger1204.5735, Elben2203.11374, Notarnicola2112.11046}, particularly through the \emph{classical shadow tomography} \cite{Huang2002.08953,2021arXiv211002965L,2020arXiv200615788H,PRXQuantum.3.020365,PhysRevLett.127.110504,2022arXiv220808964H,2022arXiv220307263H,2022arXiv220307309S,hu2022efficient,PRXQuantum.2.030348,PhysRevLett.126.050501,Koh2022classicalshadows}.

Classical shadow tomography converts quantum states to classical data with a superior sample efficiency. It can estimate expectation values of $M$ observables using only $\sim(\log M)\Vert O\Vert_\text{shd}^2$ independent randomized measurements \cite{Huang2002.08953,Paini1910.10543}, saturating the theoretical optimal bound on sample efficiency. Nevertheless, the constant coefficient $\Vert O\Vert_\text{shd}^2$, known as the \emph{operator shadow norm}, does depend on the type of observable $O$ and the randomized measurement scheme. For example, the single-qubit (local) Pauli measurements are efficient for predicting local observables, while the global Clifford measurements are efficient in estimating certain global properties such as quantum fidelity. However, an efficient and scalable approach to interpolate the local and global limits is still missing in classical shadow tomography.

\begin{figure}
    \centering
    \includegraphics[scale=0.64]{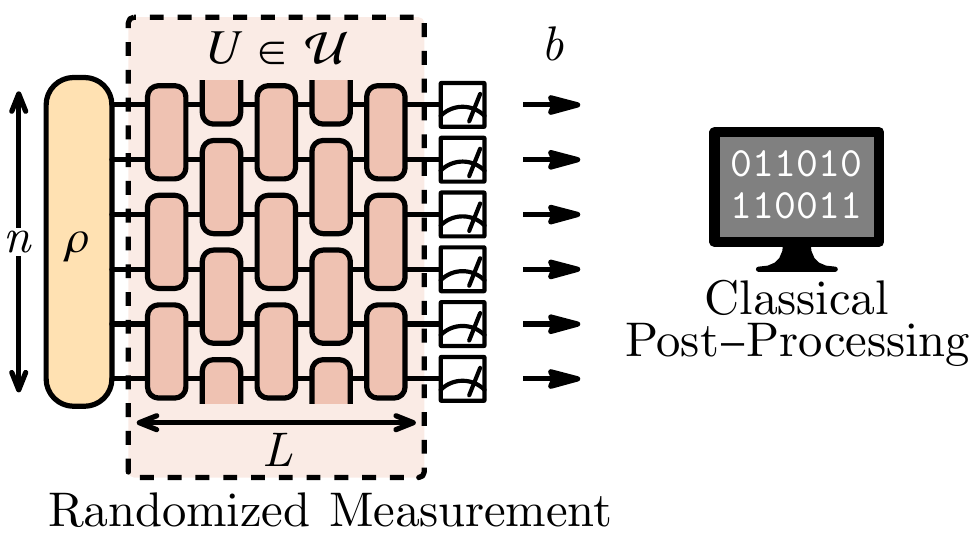}
    \caption{Classical shadow tomography in a system of $n$ qubits by randomized measurement via a finite-depth local random unitary circuit of $L$ layers. 
    }
    \label{fig:CST}
\end{figure}

As illustrated in \figref{fig:CST}, the randomized measurement protocol in classical shadow tomography is generally realized by first transforming the quantum state $\rho$ of interest by a random unitary circuit $U$ sampled from a random unitary ensemble $\mathcal{U}$, then performing a computational basis measurement and collecting the measurement outcomes $b$ (as a bit-string). Much of the literature has focused on two measurement protocols: the single-qubit Pauli measurements (where $\mathcal{U}=\mathrm{Cl}(2)^{\otimes n}$) and the global Clifford measurements (where $\mathcal{U}=\mathrm{Cl}(2^n)$). In terms of the depth $L$ of the random unitary circuit $U\in\mathcal{U}$, we may think of these two randomized measurement schemes as the $L\rightarrow 0$ and $L\rightarrow \infty$ limits of randomized measurement protocols defined on more general finite-depth unitary ensembles  (see \figref{fig:CST}). This work aims to explore the intermediate measurement schemes based on finite-depth circuits between these two limits. For simplicity, we focus on the collection of unitary ensembles defined by the brick wall arrangement of two-local random unitary gates as shown in \figref{fig:CST}, but our approach is directly applicable to other circuit structures as well. 

The ability to adjust the circuit depth $L$ (or even the circuit structure) provides great flexibility to modify the measurement scheme adaptively so as to optimize the tomography efficiency for the given target observables. This can be particularly useful when the target observables are not neatly local or low-rank (such that neither Pauli nor Clifford measurement is optimal). The shallow circuit implementation of randomized measurements is also friendly to near-term quantum devices. 
Motivated by these objectives, \refcite{Hu2107.04817} develops the foundation for shallow-circuit classical shadow tomography based on the theory of \emph{locally scrambled} quantum dynamics and the \emph{entanglement feature} formalism \cite{You1709.01223, You1803.10425, Kuo1910.11351, Akhtar2006.08797, Fan2002.12385, Hu2102.10132}. The key finding is that for a large class of randomized measurements, the classical shadow reconstruction map only depends on the entanglement properties of the classical snapshot states $U^\dagger \ket{b}$, which can be computed by solving a quantum entanglement dynamics problem.

\begin{figure}
    \centering
    \includegraphics[scale=0.64]{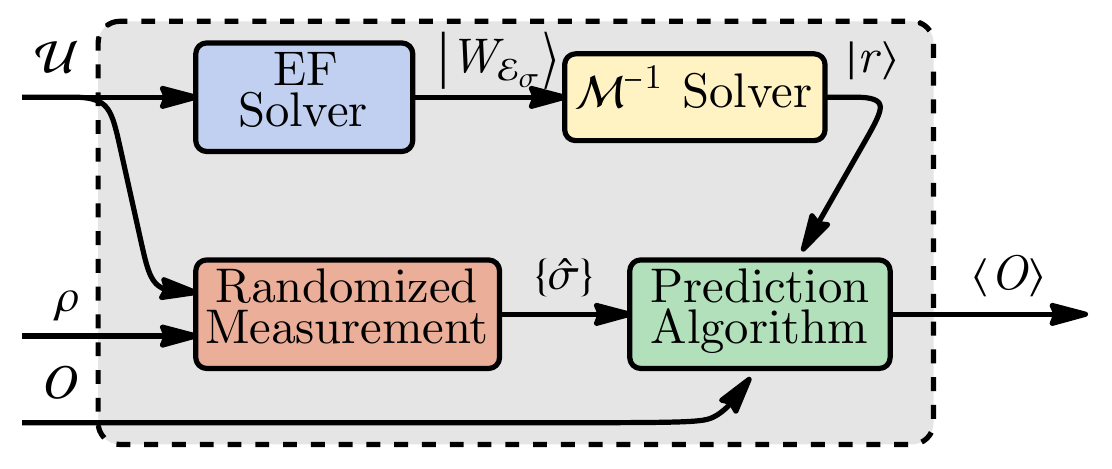}
    \caption{We outline the procedure for predicting $\langle O \rangle = \Tr(\rho O)$ using the shallow circuit classical shadow tomography approach. The ingredients are the state $\rho$, the observable $O$, and the unitary ensemble $\mathcal{U}$ defining the measurement protocol. The entanglement feature $W_{\mathcal{E}_\sigma}$ is totally determined by the second moment of the unitary ensemble $\mathcal{U}$. The snapshot states $\hat\sigma$ is collected from randomized measurements given $\mathcal{U}$ and $\rho$. The reconstruction map $\mathcal{M}^{-1}$ is determined by the entanglement features only, and is encapsulated by the reconstruction coefficients $r$. These reconstruction coefficients $r$, alongside the snapshot states $\hat\sigma$ and the observable $O$ are used to make predictions. Importantly, the entire algorithm only involves matrix product states and stabilizer states. Both are efficient representations of quantum states that can be processed on classical computers.}
\label{fig:StepsRCM}
\end{figure}

However, the reconstruction algorithm demonstrated in \refcite{Hu2107.04817} is based on a brute-force computation, which is not scalable to larger systems (the complexity for classical post-processing will be exponential in system size). The largest system size achieved in \refcite{Hu2107.04817} was nine qubits. Facing this challenge, in this work, we further develop efficient numerical methods based on \emph{matrix product state} (MPS) \cite{Orus1306.2164, Cirac2011.12127,  Geng_2022, 10.21468/SciPostPhys.10.2.040} for calculating entanglement features and solving reconstruction coefficients. The key idea is to pack the entanglement feature (purities in different entanglement regions) of the classical snapshot state $U^\dagger \ket{b}$ into a fictitious quantum many-body state and represent this entanglement feature state by an MPS. This enables efficient computation of the entanglement dynamics and finding the classical shadow reconstruction map. Our tomography procedure is outlined in \figref{fig:StepsRCM}. Using this approach, we can perform the classical shadow post-processing on 22 qubits ($n=22$) for the randomized measurement scheme based on three-layer random Clifford circuits ($L=3$) with superior sample efficiency. The primary advantages of this approach are its scalability and flexibility, i.e.~the classical post-processing can be performed for large system sizes with \emph{polynomial complexity} (given a fixed amount of measurements), while the method applies to a wide range of random Clifford measurement schemes.

The paper is organized as follows. In \secref{background}, we review the basic setup of classical shadow tomography and the key concepts of entanglement feature formalism. In \secref{algorithms}, we describe the general algorithm/procedure for computing the expectation values of generic observables and analyzing the corresponding shadow norm. In \secref{numerics}, we demonstrate the shallow-circuit classical shadow tomography on the Greenberger-Horne-Zeilinger (GHZ) state and the cluster state. The simulation confirms an unbiased tomographic reconstruction with favorable sample efficiency compared to that of Pauli measurements. In \secref{conclusion}, we suggest some future lines of inquiry and summarize the key results.

\section{Background}\label{background}

\subsection{General Framework of Classical Shadow Tomograhy}

Classical shadow tomography is an efficient protocol for predicting features of a quantum state based on a few measurements \cite{Huang2002.08953}. The trick is to use entanglement generated from a random unitary ensemble $\mathcal{U}$ to form approximate classical ``shadows'' $\hat{\rho}$ of the original state $\rho$ which are then used for predicting the desired observable. Different unitary ensembles produce different collections of shadows, which are suited for predicting different kinds of observables.

The protocol consists of two steps: randomized measurement and classical post-processing, as illustrated in \figref{fig:CST}. Starting with the initial state $\rho$, apply a randomly sampled unitary $U\in\mathcal{U}$ so that $\rho\rightarrow U\rho U^\dagger$. Then perform a projective measurement on the transformed state $U\rho U^\dagger$ in the computational basis ($Z$-basis on each qubit independently). The resulting bit-string state $\ket{b}$ (labeled by the bit-string $b$ of measurement outcomes), as well as the performed unitary transformation $U$, will be recorded. These two pieces of information define a \textit{snapshot state} $\hat{\sigma}= U^\dagger \ket{b}\bra{b} U$. The randomized measurement protocol can be formulated as a quantum channel $\mathcal{M}$, which maps the initial state $\rho$ to the average snapshot state $\sigma$ by
\begin{equation}\label{eq: def M}
\sigma=\mathcal{M}(\rho):=\mathop{\mathbb{E}}_{\hat{\sigma}\in\scE_{\sigma|\rho}} \hat{\sigma}=2^n\mathop{\mathbb{E}}_{\hat{\sigma}\in\scE_\sigma} \hat{\sigma}\Tr(\hat{\sigma}\rho),
\end{equation}
where $\mathcal{E}_\sigma=\{U^\dagger\ket{b}\bra{b}U\;|\;b\in\{0,1\}^{\times n}, U\in\mathcal{U}\}$ denotes the \emph{prior} ensemble of all possible classical snapshots. In the prior snapshot ensemble, the joint probability to sample a $(b,U)$ pair is $P(b,U)=2^{-n}P(U)$, where $P(U)$ is the probability to sample $U$ and $2^{-n}$ is the probability to sample $b$ uniformly (independent of $\rho$). Given the observation of $\rho$, the \emph{posterior} snapshot ensemble $\scE_{\sigma|\rho}$ is defined by deforming the joint probability distribution to $P_\rho(b,U)=\bra{b}U\rho U^\dagger\ket{b}P(U)$. The measurement channel $\mathcal{M}$ can be formulated based on either of the snapshot ensembles equivalently, as stated in \eqnref{eq: def M}.

Furthermore, if this measurement channel $\mathcal{M}$ is tomographically complete, i.e.~distinct states $\rho$ can be distinguished by different measurement outcomes $(b,U)$, then there is an inverse map $\mathcal{M}^{-1}$ called the \textit{reconstruction map} such that $\rho=\mathcal{M}^{-1}(\sigma)$. The image of a given snapshot state $\hat{\sigma}$ under the reconstruction map is a \textit{classical shadow} $\hat{\rho}=\mathcal{M}^{-1}(\hat{\sigma})$, which provides a single-shot classical representation of the initial quantum state $\rho$. The initial state $\rho$ can be restored as the ensemble expectation of classical shadows $\rho=\mathbb{E}_{\hat{\sigma}\in\scE_{\sigma|\rho}}\mathcal{M}^{-1}(\hat{\sigma})$, which enables us to predict  physical observables of the initial state based on the classical snapshots collected from randomized measurements,
\begin{equation}
    \langle O \rangle = \Tr(O\rho) 
    = \mathop{\mathbb{E}}_{\hat{\sigma}\in\scE_{\sigma|\rho}}\Tr(O\mathcal{M}^{-1}(\hat{\sigma})).
\end{equation}
This computation will be carried out on a classical computer. It will be efficient if the snapshot states are stabilizer states. Since the reconstruction map $\mathcal{M}^{-1}$ is not a quantum channel (because it is not a positivity-preserving map), the resulting shadow states $\hat{\rho}$ may not be positive semi-definite. Therefore, one may encounter unphysical single-shot estimation values of $O$ for certain individual samples. Nevertheless, the ensemble average of all single-shot estimations is physical and unbiased. 

In practice, quantum devices can have a significant amount of noise which can introduce randomness into meaurement outcomes. One solution to deal with noise is to incorporate the noise into the measurement channel itself. The idea is that the prior snapshot ensemble is composed of several layers of channels acting on an initial product state $\mathcal{E}_\sigma=\mathcal{E}_1 \circ \mathcal{E}_2 \dots \mathcal{E}_L (\ket{b}\bra{b}) $ which are averaged over to form the measurement channel. Incorporating noise amounts to simply introducing new noise layers resulting in a new measurement channel $\mathcal{M}'$ based on the statistical properties of the noise. If we assume that the noise is local and does not have a preferential basis, then the approach outlined in this paper would still be applicable and the resulting reconstruction map $(\mathcal{M}')^{-1}$ should have an efficient MPO description in the shallow circuit limit. 

For any finite-sized classical shadow ensemble, the estimation exhibits statistical fluctuations around the true expectation value $\langle O\rangle$. The variance in the single-shot estimation of any observable $O$ can be quantified through its \textit{operator shadow norm} $\|O\|_\text{shd}^2$, introduced in \refcite{Huang2002.08953}. The estimated expectation value obtained from averaging over $M$ samples will be within the error bound of $\|O\|_\text{shd}^2/M$ with high probability. For a given class of observables, it is possible to choose the random unitary ensemble $\mathcal{U}$ wisely so as to reduce the shadow norm and to optimize the sample efficiency \cite{Hu2107.04817}.

\subsection{Locally-Scrambled Classical Shadow Tomography}

Restricting the randomized measurement scheme to the scope of locally-scrambled ensembles proves to be advantageous \cite{Hu2107.04817}. Locally-scrambled ensembles \cite{Kuo1910.11351} are random unitary ensembles which are invariant under both left and right local basis transformations, i.e.~the probability distribution for all $U\in\mathcal{U}$ satisfies $P(U)=P(UV)=P(VU)$ where $V=\bigotimes_i V_i$ with $V_i \in \mathrm{U}(2)$ being any local basis transformation. Many typically studied random circuit models are in fact, local basis invariant, such as random Haar circuits \cite{Nahum1608.06950, Nahum1705.08975, Zhou1804.09737, Bao1908.04305, Choi1903.05124, Fan2002.12385} and quantum Brownian circuits \cite{Lashkari1111.6580, Gharibyan1803.08050, Zhou1805.09307, Xu1805.05376, Chen1808.09812}. Furthermore, since the reconstruction map only depends on the second moment of the unitary ensemble, Clifford circuits, which are typically used for classical shadow tomography because of their classical simulability, are equivalent in terms of their tomographic properties to the random Haar circuit. Therefore, the formulation of locally-scrambled classical shadow tomography in \refcite{Hu2107.04817} is applicable to generic random Clifford circuits of any depth and any gate arrangement.

In the locally-scrambled case, it was shown that the reconstruction map is determined entirely by the average purity (exponential of 2nd R\'enyi entropy) of the (prior) classical snapshot states \cite{Hu2107.04817} in \emph{all} possible sub-regions, which are also known as the \textit{entanglement features} \cite{You1803.10425,Kuo1910.11351,Fan2002.12385,Akhtar2006.08797}. One might be daunted by the fact that there are exponentially many entanglement features to keep track of, one for every possible sub-region of the system. Thus the classical post-processing will not be scalable. This paper aims to answer the challenge. In particular, two questions will be addressed:
\begin{enumerate}
\item Is there any hope of representing the entanglement features efficiently? 
\item Furthermore, can we utilize such representation to determine and apply the reconstruction map efficiently to large systems? 
\end{enumerate}

The answer to the first question is yes: average sub-region purities of locally-scrambled systems have efficient representations as matrix product states (MPS) due to the fact that purities are completely positive and many-body states of positive wave functions are typically area-law entangled \cite{Grover_2015}, such that the MPS bond dimension is a constant that does not scale with the system size $n$. The key idea is that the average purity of a $n$-qubit quantum system over all sub-regions can be compactly encoded in a fictitious many-body state vector, called the \emph{entanglement feature state}. Given an ensemble $\mathcal{E}$ of random states (random density matrices) of the system, the entanglement feature state $\ket{W_\mathcal{E}}$ associated with $\mathcal{E}$ is a vector whose components are 
\begin{equation}\label{defEFstate}
(W_\mathcal{E})_A=\braket{A}{W_\mathcal{E}}:=\mathop{\mathds{E}}_{\rho\in\mathcal{E}}\Tr_A(\Tr_{\bar{A}}\rho)^2,
\end{equation}
where $A$ denotes a sub-region of the system (as a subset of qubits) and $\bar{A}$ is the complement region of $A$. For each state $\rho$ in the ensemble $\mathcal{E}$, $\rho_A=\Tr_{\bar{A}}\rho$ is the reduced density matrix of $\rho$ in the region $A$, by tracing out qubits in the region $\bar{A}$. The entanglement feature $(W_\mathcal{E})_A$ in the region $A$ is simply the purity of $\rho_A$ averaged over all random states $\rho$ in the ensemble $\mathcal{E}$. The entanglement feature state $\ket{W_\mathcal{E}}$ is a superposition of the basis states $\ket{A}$ labeled by sub-regions, and the superposition coefficients are the average purities in the corresponding sub-regions. Note that the entanglement feature state is not a physical state of the quantum system. It merely encodes the entanglement properties of the underlying physical quantum states. For example, the ensemble of \emph{pure product states} has the following entanglement feature state
\begin{equation}\label{Wprod}
    \ket{W_\text{prod}}=\sum_{A}\ket{A},
\end{equation} because the purity is unity in any sub-region of any pure product state. In this way, we can efficiently encode the purity information over all sub-regions in a single entanglement feature state $\ket{W_\mathcal{E}}$.

The advantage of the entanglement feature formalism lies in the fact that there exists an efficient numerical method to calculate the \emph{time evolution} of the entanglement feature state as the underlying quantum state evolves in time together. Let $\mathcal{U}$ be a locally-scrambled random unitary ensemble and $\mathcal{E}$ be a random state ensemble. One can define a new random state ensemble $\mathcal{E}'=\{U\rho U^\dagger\;|\;\rho\in\mathcal{E}, U\in\mathcal{U}\}$, which can be interpreted as the ensemble of the random states evolved by the random unitaries. Then the entanglement features of $\mathcal{E}'$ and $\mathcal{E}$ are related by
\begin{equation}\label{EFdynamics}
\ket{W_{\mathcal{E}'}}=\hat{W}_{\mathcal{U}}\hat{W}_{\id}^{-1}\ket{W_{\mathcal{E}}},
\end{equation}
where $\hat{W}_{\mathcal{U}}$ is called the \emph{entanglement feature operator} associated with the random unitary ensemble $\mathcal{U}$, whose matrix components are given by $\bra{A}\hat{W}_{\mathcal{U}}\ket{B}=\mathop{\mathds{E}}_{U\in\mathcal{U}}\Tr_{A,B}(\Tr_{\bar{A},\bar{B}}U)^2$ with $A$ and $B$ being the sub-regions on the future and the past sides of $U$ respectively. $\hat{W}_{\id}$ is the entanglement feature operator for the identity operator, and its inverse operator is denoted as $\hat{W}_{\id}^{-1}$. Therefore, each step of the unitary evolution of the physical state ensemble can be mapped to a step of transfer matrix evolution of the entanglement feature state.

An efficient numerical approach to compute the entanglement feature dynamics \eqnref{EFdynamics} has been developed  \cite{Akhtar2006.08797, Fan2002.12385} based on the MPS representation of entanglement feature states.
\begin{equation}\label{ef-mps-eq}
(W_\mathcal{E})_A=\braket{A}{W_\mathcal{E}}=\Tr \prod_i (\delta_{i\notin A}\mathsf{W}_{\mathcal{E},i}^0+\delta_{i\in A}\mathsf{W}_{\mathcal{E},i}^1).
\end{equation}
Here $\mathsf{W}_{\mathcal{E},i}^0$ and $\mathsf{W}_{\mathcal{E},i}^1$ are two sets of matrices parametrizing the MPS, where $i$ labels the qubit/site. If the state ensemble $\mathcal{E}$ is translation invariant, we may choose to drop the index $i$ on the MPS matrices $\mathsf{W}_{\mathcal{E},i}^0,\mathsf{W}_{\mathcal{E},i}^1$. Furthermore, for the pure state entanglement feature, the MPS matrices must satisfy various constraints. For example, for pure states, the purity of the whole system is one i.e. $\Tr(\mathsf{W}_{\mathcal{E}}^0)^n=\Tr(\mathsf{W}_{\mathcal{E}}^1)^n=1$. In addition, because the purity of complementary regions is the same, i.e.~$(W_{\mathcal{E}})_A=(W_{\mathcal{E}})_{\bar{A}}$, the entanglement feature state has a $\mathbb{Z}_2:A\to\bar{A}$ symmetry, so the MPS tensor (formed by combining the two matrices into a single tensor) must carry definite $\mathbb{Z}_2$ representation. These symmetries greatly constrain the form of the MPS matrices (see \refcite{Akhtar2006.08797} for more discussions).

As a fictitious many-body state, the entanglement feature state is typically area-law entangled due to its completely positive sign structure \cite{Grover_2015}. Thus it generally admits efficient MPS representations if the system is one-dimensional. For example, the entanglement feature state of pure product states \eqnref{Wprod} can be represented as a trivial MPS (a product state) with bond dimension one. Most typical entanglement feature states will have a low bond dimension even as the underlying physical state ensemble is of volume-law entanglement. For nearly all locally-scrambled dynamics with various entanglement production rates, it has been numerically observed \cite{Akhtar2006.08797} that a bond dimension $D_W=2$ MPS state could capture the purity evolution in all sub-regions to high accuracy. The accuracy can be further improved with a larger bond dimension. The evolution of the entanglement feature state, for example, by applying alternating layers of brick wall unitaries, can also be performed at the MPS level using the entanglement feature transfer matrix following \eqnref{EFdynamics}. To implement that, we use a time-evolved block decimation (TEBD) approach and truncate at a fixed bond dimension $D_W$ (see Appendix C of \refcite{Fan2002.12385} or Appendix A, C of \refcite{Akhtar2006.08797} for further details). With these previously developed tools, one can efficiently calculate the evolution of the average purity given a unitary ensemble. Furthermore, given the same circuit geometry, all unitary two-designs have the same entanglement feature dynamics, and so the Clifford circuit entanglement feature transfer matrix is equivalent to the random Haar one.

For the purpose of classical shadow tomography, what we need to know is the entanglement feature of the classical snapshots. Given any randomized measurement scheme (as specified by the random unitary ensemble $\mathcal{U}$), all possible classical snapshots form a random state ensemble, denoted as
\begin{equation}\label{eq: snapshot ensemble}
\mathcal{E}_\sigma=\{U^\dagger\ket{b}\bra{b}U\;|\;b\in\{0,1\}^{\times n}, U\in\mathcal{U}\}.
\end{equation}
This is also the prior snapshot ensemble defined below \eqnref{eq: def M}. Following the definition in \eqnref{defEFstate}, we can define the entanglement feature state $\ket{W_{\mathcal{E}_\sigma}}$ for the classical snapshot ensemble $\mathcal{E}_\sigma$, whose components are \begin{equation}
    (W_{\mathcal{E}_\sigma})_A=\braket{A}{W_{\mathcal{E}_\sigma}}:=\mathop{\mathds{E}}_{\hat{\sigma}\in\mathcal{E}_\sigma}\Tr_A(\Tr_{\bar{A}}\hat{\sigma})^2.
\end{equation} Note that the classical snapshot state $\hat{\sigma}=U^\dagger\ket{b}\bra{b}U$ is always constructed from a (reversed) unitary evolution $U^\dagger$ on a pure product state $\ket{b}$. So to obtain $\ket{W_{\mathcal{E}_\sigma}}$, we only need to take the MPS representation of $\ket{W_\text{prod}}$ in \eqnref{Wprod} as the initial state and use the TEBD algorithm to evolve the entanglement feature MPS step by step following the entanglement feature dynamics in  \eqnref{EFdynamics}, as the underlying snapshot state gets evolved by the (reversed) Clifford circuit $U^\dagger$ layer by layer. Computing the entanglement feature $\ket{W_{\mathcal{E}_\sigma}}$ from the random unitary ensemble $\mathcal{U}$ is denoted as ``EF Solver'' in \figref{fig:StepsRCM}. The algorithm complexity is linear in the circuit depth $L$ and independent of the number of qubit $n$ (assuming translation symmetry).

According to \refcite{Hu2107.04817}, the entanglement feature of classical snapshots $\ket{W_{\mathcal{E}_\sigma}}$ is all we need to obtain the classical shadow reconstruction map $\mathcal{M}^{-1}$ for the corresponding randomized measurement scheme, as long as the random unitary ensemble $\mathcal{U}$ is locally scrambled (which applies to random Clifford circuits). The steps to obtain $\mathcal{M}^{-1}$ are as follows:

\begin{figure}[h]
\includegraphics[scale=0.7]{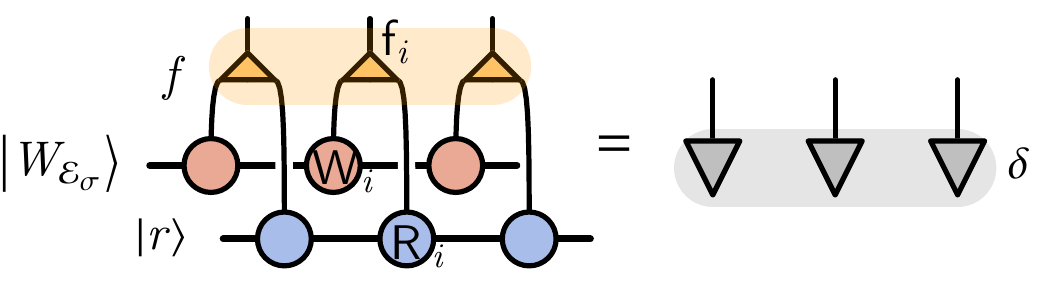}
\caption{Illustration of the MPS equation \eqnref{consistency-eq}. The fusion tensor $\mathsf{f}_i$ is defined in \tabref{tab: f_tensor}. The reconstruction coefficient $\ket{r}$ can be determined from the entanglement feature $\ket{W_{\mathcal{E}_\sigma}}$ by solving this MPS equation.}
\label{fig: MPS}
\end{figure}
First, one solves for the \textit{reconstruction coefficients} $\ket{r}$ (or $r_A=\braket{A}{r}$) from $\ket{W_{\mathcal{E}_\sigma}}$ through the following consistency equation derived in \refcite{Hu2107.04817} (which is essentially requiring $\mathcal{M}^{-1}\mathcal{M}=\id$ at the quantum channel level)
\begin{equation}\label{consistency-eq}
    \sum_{A,C} r_A f_{A,B,C} (W_{\mathcal{E}_\sigma})_C = \delta_{B=\Omega}.
\end{equation}
Here, $A,C$ are summed over all sub-regions of the system (as subsets of qubits). $\Omega$ stands for the full system (as the full set of qubits) and $\delta_{B}=1$ if $B=\Omega$ otherwise, $\delta_{B}=0$. Also, $r_A$ and $(W_{\mathcal{E}_\sigma})_C$ are components of the corresponding vectors $\ket{r}$ and $\ket{W_{\mathcal{E}_\sigma}}$, and $f_{A,B,C}$ is a fusion tensor that can be factorized to each site as $f=\bigotimes_{i=1}^{n} \mathsf{f}_i$, where components of the on-site tensor $\mathsf{f}_i$ are given by \tabref{tab: f_tensor}.

\begin{table}[h]
    \centering
    \begin{tabular}{c|cccc}
    \multirow{2}{5pt}{$\mathsf{f}_i$} & \multicolumn{2}{c}{$i\notin A$} & \multicolumn{2}{c}{$i\in A$} \\
    & $i\notin C$ & $i\in C$ & $i\notin C$ & $i\in C$\\
    \hline
    $i\notin B$ & 2 & 0 & $\frac{8}{3}$ & $-\frac{4}{3}$\\
    $i\in B$ & 0 & 0 & $-\frac{2}{3}$ & $\frac{4}{3}$ \\
    \end{tabular}
    \caption{Components of the $\mathsf{f}_i$ tensor (for qubit systems). The systematic formula for $\mathsf{f}_i$ in generic qudit systems can be found in \refcite{Hu2107.04817}.}
    \label{tab: f_tensor}
\end{table}

Using MPS representations for both $\ket{r}$ and $\ket{W_{\mathcal{E}_\sigma}}$, and the tensor product structure of $f$, the left hand side of \eqnref{consistency-eq} may be represented as an MPS equation, as shown in \figref{fig: MPS}. The left-hand side of \eqnref{consistency-eq} is made up of an MPS of bond dimension $D_r \times D_W$, whereas the right-hand side of \eqnref{consistency-eq} is a trivial MPS (a product state). We may then solve for the MPS tensors describing $\ket{r}$ by minimizing the following loss function
\begin{equation}\label{eq: loss}
\begin{split}
    \mathcal{L}&=\Vert\mathcal{M}^{-1}\mathcal{M}-\id\Vert^2\\
    &=\sum_{B}\Big(\sum_{A,C} r_A f_{A,B,C} (W_{\mathcal{E}_\sigma})_C - \delta_{B,\Omega}\Big)^2.
\end{split}
\end{equation}
The optimization is achieved by the gradient descent algorithm using the PyTorch package \cite{NEURIPS2019_9015} for auto-differentiation. The loss function can be suppressed to $\mathcal{L}\lesssim 10^{-3}$ 
with a modest bond dimension $D_r= 6$ for circuit depths $L\leq 5$. In summary, for shallow circuits, we numerically observe that the reconstruction coefficients can be well described by a low bond-dimension MPS. Solving the reconstruction coefficient $\ket{r}$ from the entanglement feature $\ket{W_{\mathcal{E}_\sigma}}$ is denoted as the ``$\mathcal{M}^{-1}$ Solver'' in \figref{fig:StepsRCM}. The algorithm complexity is \emph{linear} in the number $n$ of qubits and \emph{independent} of the circuit depth $L$.

Then, the reconstruction map is a linear functional of the reconstruction coefficients. 
\begin{equation}\label{rcmap}
\mathcal{M}^{-1}(\sigma)=2^{n}\sum_{A} r_A \sigma_A 
\end{equation}
Each reconstruction coefficient is weighted by a partial trace of the input $\sigma$ embedded back into the Hilbert space so that the sum is over normalized density matrices on the full Hilbert space. More specifically, $\sigma_A = 2^{|A|-n} (\Tr_{\bar{A}}\sigma) \otimes \id^{\otimes (n-|A|)} $, which is a linear functional on $\sigma$. In practice, one works with individual snapshot states as opposed to their ensemble average. For Clifford circuit-based randomized measurement, each individual snapshot state is a stabilizer state, which admits efficient post-processing on a classical computer. 

We would like to mention that the analytic expression for the reconstruction coefficient $\ket{r}$ is known from \cite{Bu2022} by solving \eqnref{consistency-eq}
\begin{equation}\label{r-to-ef}
r_A = 2^{-n}(-1)^{|A|}\sum_{C|C\supseteq A}\frac{3^{|C|}}{\sum_{B |B\subseteq C} (-2)^{|B|}(W_{\mathcal{E}_\sigma})_B}.
\end{equation}
Nevertheless, it does not seem to help our purpose, as we do not know how to encode this expression as a MPS efficiently. Directly evaluating \eqnref{r-to-ef} for every sub-region $A$ is not scalable. Therefore, we will still follow our approach to solve \eqnref{r-to-ef} by MPS optimization.

In the next section, we will answer the second question: how can we leverage the scalability and efficiency of matrix product states for prediction using classical shadow tomography?

\section{Applications for scalable prediction}\label{algorithms}

\subsection{Matrix Product State Representation of Reconstruction Coefficients}

We consider that the quantum system is made of $n$ qubits arranged on a one-dimensional lattice. We assume that there exists an efficient MPS representation for the reconstruction coefficients $r_A$. The existence of an efficient representation for $\ket{r}$ is implied from the tensor network structure of the self-consistency equation and the efficient MPS representation for the entanglement feature. Let $\mathsf{R}_i^j, j=0,1,i\in\{1\cdots n\}$ be the MPS matrices for $\ket{r}$ at site $i$ of type $j$, i.e. $\mathsf{R}_i^j$ is a $D_r\times D_r$ real matrix, satisfying 
\begin{equation}\label{r-mps-eq}
r_A=\braket{A}{r}=\Tr\left(\mathsf{B}\prod_{i=1}^{n} (\delta_{i\notin A}\mathsf{R}_{i}^0+\delta_{i\in A}\mathsf{R}_{i}^1) \right),
\end{equation}
where $\mathsf{B}$ is a $D_r\times D_r$ real matrix serving as a twisted periodic boundary condition for the MPS. Here the site index $i=1,2,\cdots,n$ labels the qubits on the one-dimensional lattice. For $L=0,\infty$, we may drop the site index altogether because there is translation invariance. For $0<L<\infty$, there is only a two-site translation invariance, so only the parity of the site is important. For example, consider the deep circuit limit ($L=\infty$), which corresponds to random Clifford measurements. In this case, $r_{\emptyset}=-1$ and $r_{\Omega}=1+2^{-n}$, where $2$ is the local Hilbert space dimension and $n$ is the number of qubits. One choice of tensors here is
\begin{equation}\label{clifford-r}
    \mathsf{R}^j = \begin{cases} \begin{pmatrix} 1 & 0 \\ 0 & 0 \end{pmatrix} & j=0 \\ \begin{pmatrix} 0 & 0 \\ 0 & r_\Omega^{1/n} \end{pmatrix} & j=1 \end{cases} \quad \mathsf{B} = \begin{pmatrix} -1 & 0 \\ 0 & 1 \end{pmatrix}.
\end{equation}
As the MPS matrix $\mathsf{R}_{i}^{j}$ is translationally invariant, we will drop the site index $i$ and denote it as $\mathsf{R}^{j}$. 
Introducing the boundary tensor can help simplify calculations. For example, without inserting the boundary tensor (or equivalently setting $\mathsf{B}=\id$), we have that $\lambda_1^n + \lambda_2^n = -1$ for the $\emptyset$ component, where $\lambda_1,\lambda_2$ are the eigenvalues of $\mathsf{R}^0$. When $n$ is even, the equation has no real solutions, so the eigenvalues must be complex. In practice, this makes solving for the $\mathsf{R}$ matrices more difficult, as one has to optimize over complex tensors; in addition, one has to perform complex tensor network contractions, which can also significantly slow down calculations. This can be avoided, as suggested above, through a boundary tensor. 

On the other hand, in the shallow circuit limit $L= 0$, the circuit reduces to random Pauli measurements. In this case, a trivial product state MPS suffices with 
\begin{equation}\label{pauli-r}
    \mathsf{R}^{j}=\begin{cases}-1 & j=0 \\ 3/2 & j=1 \end{cases}
\end{equation}
and $\mathsf{B}=1$. Note that translation invariance is restored is both limits because $\mathsf{R}^0, \mathsf{R}^1$ does not depend on the site index $i$ explicitly. For finite-depth circuits, $0<L<\infty$, the brick wall structure breaks the translation invariance into the subgroup of two-site translations. 

Also, as expected, the MPS representation of $\ket{r}$ both the shallow and deep circuit limits have low bond dimensions, although the underlying snapshot states have very different entanglement scalings (from area-law in $L=0$ to volume-law in $L\to \infty$). The MPS representation of the reconstruction coefficients $r_A$ allows us to apply the reconstruction map \eqnref{rcmap} in prediction algorithms in a scalable way, which we will see next. Even simply knowing the reconstruction coefficients exactly does not solve this problem because there are an extensive number of them. Therefore, the MPS representation of the reconstruction coefficients is key for classical shadow tomography (beyond the Pauli measurement) to be feasible on near-term quantum devices with large numbers of qubits; otherwise, we cannot reconstruct classical shadows efficiently. 

\begin{figure}[htbp]
\begin{center}
\includegraphics[scale=0.7]{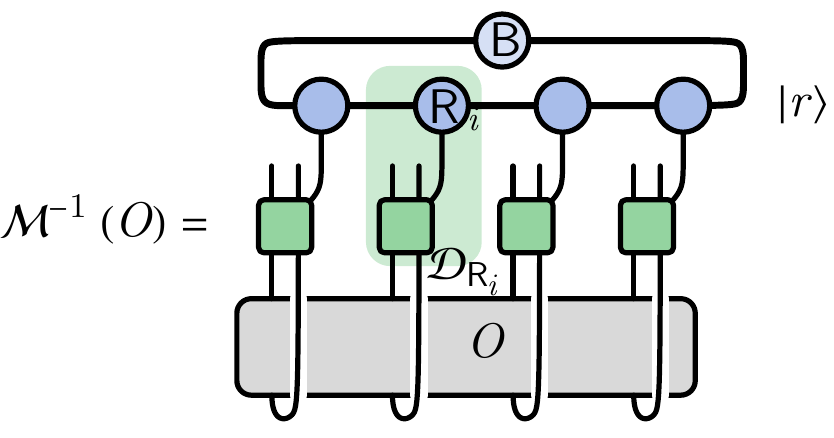}
\caption{Tensor network representation of the reconstruction map $\mathcal{M}^{-1}(O)$ acting on a generic operator $O$, as defined in \eqnref{rcm}. Thick lines represent virtual bonds in the MPS. Each green box represents a dephasing channel $\mathcal{D}_{\mathsf{R}_i}$ at the qubit $i$ in \eqnref{eq: def dephasing}. The dephasing channels are correlated together by the MPS to form the reconstruction map.}
\label{fig:reconstruction}
\end{center}
\end{figure}

Given the MPS representation of the reconstruction coefficient $\ket{r}$, the reconstruction map $\mathcal{M}^{-1}$ may be viewed as a matrix product operator (MPO). The local tensors are a generalized dephasing channel $\mathcal{D}_{\mathsf{R}_i}$ defined in terms of the MPS tensor for the reconstruction coefficients at a particular site $i$. It acts on a generic operator $O$ as
\begin{equation}\label{eq: def dephasing}
    \mathcal{D}_{\mathsf{R}_i}(O) := \mathsf{R}_{i}^0 \otimes (\id \Tr_i(O)/2) + \mathsf{R}_{i}^1 \otimes O
\end{equation}
%
Then the reconstruction map on an arbitrary operator can be given as a super-operator defined in terms of the MPS tensors for the reconstruction coefficients, see \figref{fig:reconstruction},
\begin{equation}\label{rcm}
    \mathcal{M}^{-1} = 2^n  \Tr_\text{virtual} \left(\mathsf{B}\prod_{i=1}^n \mathcal{D}_{\mathsf{R}_i}\right)  
\end{equation}
where the trace and product are over the virtual indices of the MPS. When acting on a single-qubit Pauli operator $P_i$, where $P_i\in\{\id,X,Y,Z\}$, the generalized dephasing channel evaluates to
\begin{equation}
    \mathcal{D}_{\mathsf{R}_i} (P_i) = (\delta_{P_i=\id} \mathsf{R}_{i}^0 + \mathsf{R}_{i}^1) \otimes P_i.
\end{equation}
Hence, single-qubit Pauli operators are eigen-operators of the generalized dephasing channel, and Pauli strings are eigen-operators of the reconstruction map. On a arbitrary $n$-qubit Pauli string $P = \bigotimes_{i=1}^n P_i$, where $P_i\in\{\id,X,Y,Z\}$ the reconstruction map evaluates to 
\begin{equation}\label{rcm-pauli}
    \mathcal{M}^{-1}(P) = 2^n \Tr \left( \mathsf{B} \prod_i ( \delta_{P_i=\id} \mathsf{R}_i^0 + \mathsf{R}_i^1 ) \right) P.
\end{equation}
The above formula is most useful for designing prediction algorithms. As we will see below, we can use the self-adjoint property of the reconstruction map to apply it to the operator whose expectation value we want to predict. This then allows us to take the expectation value of the transformed observable under the snapshot state, which has an efficient representation as a stabilizer state.

\subsection{Pauli Estimation}\label{sec: pauli}

Suppose we wish to estimate a Pauli observable $P = \bigotimes_{i=1}^n P_i$, where $P_i\in\{\id,X,Y,Z\}$. Efficient classical shadow tomography schemes for the prediction of such operators only exist using random Pauli measurements and become quickly infeasible as the weight of the Pauli string grows. Here, we show how the scheme can be extended to finite-depth circuits given the reconstruction coefficients. In practice, the expectation value is given by an average over the snapshot states $\hat{\sigma}_j$, $j=1\cdots M$.
\begin{equation}\label{eq:def_Pauli_estimate}
    \langle P \rangle = \frac{1}{M} \sum_{m=1}^M \Tr(P \mathcal{M}^{-1}(\hat{\sigma}_m)).
\end{equation}
Since the measurement channel $\mathcal{M}$ defined in \eqnref{eq: def M} is self-adjoint, so as its inverse
\begin{equation}\Tr(P \mathcal{M}^{-1}(\hat{\sigma}))=\Tr(\mathcal{M}^{-1}(P)\hat{\sigma}),
\end{equation}
we may move $\mathcal{M}^{-1}$ onto $P$, and apply \eqnref{rcm-pauli} to simplify $\mathcal{M}^{-1}(P)$, then \eqnref{eq:def_Pauli_estimate} becomes
\begin{equation}
    \langle P \rangle = 2^n \Tr \left( \mathsf{B} \prod_i ( \delta_{P_i=\id} \mathsf{R}_i^0 + \mathsf{R}_i^1 ) \right) \langle P \rangle_\sigma,
\end{equation}
where $\langle P \rangle_\sigma = \frac{1}{M}\sum_{m=1}^{M} \Tr(\hat{\sigma}_m P)$. This is a key result of this work. It enables us to extend classical shadow tomography beyond the Pauli measurement limit without losing the classical post-processing efficiency. Every term above is efficiently computable --- the trace is over a product of $n$ low-dimensional matrices and can be thought of as an inner product over MPS states, and the expectation value of the Pauli observable $P$ on a stabilizer state $\hat{\sigma}_j$ can also be computed efficiently according to the Gottesmann-Knill theorem \cite{Gottesman9807006,PhysRevA.70.052328}, which is implemented in \cite{2022arXiv220307263H}. 

\subsection{Generic Observable and Fidelity Estimation}
\label{sec: fidelity}

What about more general operators? To effectively utilize the results of the previous section for scalable prediction, we need local tensor network descriptions of operators. One approach is to view the general operator $O$ as a state in the Pauli basis i.e. 
\begin{equation}\label{pauli-component}
    (O)_P = \braket{P}{O}=\Tr(PO).
\end{equation}
The operator $O$ may be viewed as a state in the $4^n$ dimensional Hilbert space with basis states labeled by the different Pauli strings. Given this state description, we can now assign onsite matrices $\mathsf{O}_i^j$ where $i$ labels the site and the $j$ labels the Pauli at site $i$ in the matrix-product expansion of the operator. If $O$ has little operator entanglement, then the corresponding matrices $\mathsf{O}_i^j$ will also have a small bond dimension, and the decomposition can be considered efficient. 
\begin{equation}\label{mpo}
\begin{split}
    (O)_P = \Tr \Bigg(\prod_i ( &\mathsf{O}_i^0\delta_{P_i=\id} + 
    \mathsf{O}_i^1\delta_{P_i=X} \\
    &+\mathsf{O}_i^2\delta_{P_i=Y} + 
    \mathsf{O}_i^3\delta_{P_i=Z} )\Bigg).
\end{split}
\end{equation}

One context in which more general operator estimation is useful is when estimating fidelity. The fidelity $F(\rho,\rho')$ provides a way to characterize the closeness of one state $\rho$ to another $\rho'$. It is defined as $F(\rho,\rho')=(\Tr\sqrt{\sqrt{\rho}\rho'\sqrt{\rho}})^2\in[0,1]$. The greater the value of $F$, the closer the two states are. If any of the states $\rho$ or $\rho'$ is pure, their fidelity simplifies to
\begin{equation}
    F(\rho,\rho')=\Tr(\rho\rho')
\end{equation}
By viewing $F(\rho,\rho')$ as the expectation value of the operator $O=\rho'$ on the state $\rho$, the fidelity estimation amounts to predicting a low-rank, non-local observable $O=\rho'$ using the means described in this paper.

\begin{figure}[htbp]
\includegraphics[scale=0.7]{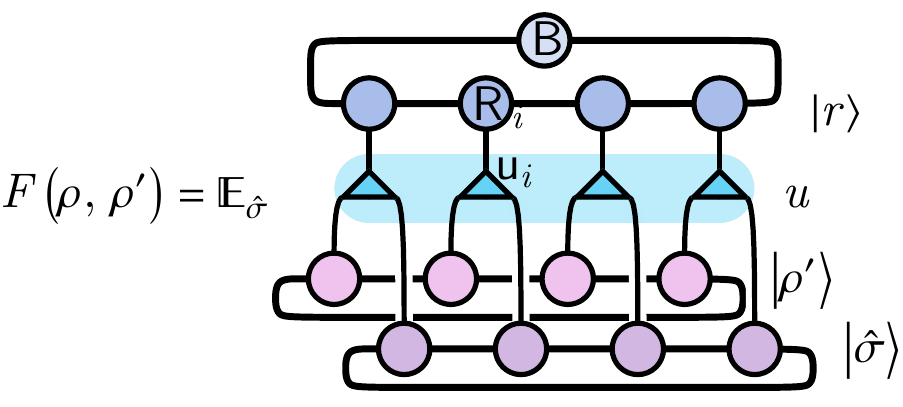}
\caption{Tensor network representation of the fidelity estimator $F(\rho,\rho')$ in \eqnref{fid-formula}. The vertex tensor  $\mathsf{u}_i$ is defined in \eqnref{eq: def u}. $\ket{\rho'}$ (or $\ket{\hat{\sigma}}$) is the vector representation of the density matrix $\rho'$ (or the snapshot state $\hat{\sigma}$) in the Pauli operator basis.}
\label{fig: fid-formula}
\end{figure}

For fidelity estimation, one might be worried that the operator entanglement of the reference state $\rho'$ is too large to represent efficiently as MPS. Though this may seem like an obstacle, ground states, low-entangled states, and stabilizer states of shallow, local circuits, all have an efficient MPS description of the form above. The last one, in particular, is useful in fidelity estimation, since we will use the MPS description of snapshot states to convert the trace into a tensor network. We will lay out the general formula given the matrix-product representation of $\rho'$ per \eqnref{mpo}, and leave details about how to construct the MPS for $\hat\sigma$ in the \appref{mps-stabs}. A straightforward application of \eqnref{rcm-pauli} gives the desired formula.
%
%
\begin{equation}\label{fid-formula}
    F(\rho,\rho')=\mathbb{E}_{\hat{\sigma}}\left[ \sum_{P} (\hat\sigma)_P (\rho')_P \Tr \left( \mathsf{B} \prod_i ( \delta_{P_i=\id} \mathsf{R}_i^0 + \mathsf{R}_i^1 ) \right) \right]
\end{equation}
Using the MPS representation for the Pauli coefficients of $\hat\sigma$ and $\rho'$, the summation above can be viewed as a tensor network, as shown in \figref{fig: fid-formula}. The idea is to take the MPS matrix for $\hat\sigma, \rho'$ and $r$ and contract them along an appropriate vertex tensor $u=\bigotimes_{i=1}^n\mathsf{u}_{i}$. If we label the indices of $\mathsf{u}_i$ as $(\mathsf{u}_i)_{P_i,P_i',j}$, where $P,P'$ are Pauli strings in the basis expansion of $\hat\sigma,\rho'$, respectively, and $j$ labels the local reconstruction MPS matrix $\mathsf{R}^j_i$, then 
\begin{equation}\label{eq: def u}
(\mathsf{u}_i)_{P_i,P_i',j} = \delta_{P_i=P_i'}(\delta_{P_i\neq \id} \delta_{j=1} + \delta_{P_i=\id} ).
\end{equation}

Note that this procedure applies also for predicting generic local operators $O$, by expanding the operator in its Pauli basis representation first. One only needs to replace $\rho'$ by $O$ in \eqnref{fid-formula}, then $\langle O \rangle=\Tr (\rho O)$ can be evaluated by a similar tensor network of the same structure in \figref{fig: fid-formula}. The procedure outlined above only relies on an efficient Pauli basis representation for the state or operator in question.

\subsection{Locally-Scrambled Shadow Norm}

Apart from estimating the expectation value of an observable $O$ from the classical shadow data, we also want to bound the variance of our estimation. This is provided by the operator shadow norm $\|O\|^2_\text{shd}$ introduced in \refcite{Huang2002.08953}. In the following, we will focus on the case that $O$ is a \emph{traceless} operator, i.e.~$\Tr O =0$. Let $\mathrm{var}\, \langle O \rangle $ be the variance of the estimation $1/M\sum_{m=1}^{M}\Tr(O\mathcal{M}^{-1}(\hat{\sigma}_m))$ obtained from averaging $M$ samples (for finite $M$). The variance of the mean decays as 
\begin{equation}\label{eq: varO shd}
    \mathrm{var}\, \langle O\rangle  \lesssim \|O\|^2_\text{shd}/M,
\end{equation}
where $\|O\|^2_\text{shd}=\max_\rho\mathbb{E}_{\hat{\sigma}\in\mathcal{E}_\sigma}2^n\Tr(\hat{\sigma}\rho)(\Tr(\hat{\sigma}\mathcal{M}^{-1}(O)))^2$ is the shadow norm. A more precise statement of \eqnref{eq: varO shd} and its proof can be found in \refcite{Huang2002.08953}. 

The conventional shadow norm $\|O\|^2_\text{shd}$ is defined as a maximum over all possible states $\rho$ of the quantum system. This definition requires information about the third moment of the snapshot state ensemble $\mathcal{E}_\sigma$, which could take some effort to compute in general. In the context of locally-scrambled ensembles, \refcite{Hu2107.04817} introduces an alternative version of the shadow norm, called the \emph{locally-scrambled shadow norm}, denoted as $\| O \|_{\mathcal{E}_\sigma}^2$. It only requires the second moment of $\mathcal{E}_\sigma$, which can be conveniently read out from the entanglement feature $\ket{W_{\mathcal{E}_\sigma}}$ computed already. The key modification in the definition is to replace the maximization $\max_\rho$ by an average of $\rho$ over its locally scrambled ensemble $\mathcal{E}_\rho=\{V\rho V^\dagger|V=\bigotimes_iV_i\in \mathrm{U}(2)^{\otimes n}\}$ with contains all states related to $\rho$ by local basis transformations $V$. The locally-scrambled shadow norm is defined as
\begin{equation}\label{eq: def shadow norm}
\begin{split}
    \| O \|_{\mathcal{E}_\sigma}^2:&=\mathop{\mathbb{E}}_{\rho\in\mathcal{E}_\rho}\mathop{\mathbb{E}}_{\hat{\sigma}\in\mathcal{E}_\sigma}2^n\Tr(\hat{\sigma}\rho)(\Tr(\hat{\sigma}\mathcal{M}^{-1}(O)))^2\\
    &=\mathop{\mathbb{E}}_{\hat{\sigma}\in\mathcal{E}_\sigma}(\Tr(\hat{\sigma}\mathcal{M}^{-1}(O)))^2,
\end{split}
\end{equation}
where the dependence on $\rho$ drops from the result due to the locally scrambled property of $\scE_\rho$. So $\| O \|_{\mathcal{E}_\sigma}^2$ is only a function of the randomized measurement scheme specified by the prior snapshot ensemble $\mathcal{E}_\sigma$ (defined in \eqnref{eq: snapshot ensemble}) and the observable $O$ in question. 

\begin{figure}[htbp]
    \centering
    \includegraphics[scale=0.7]{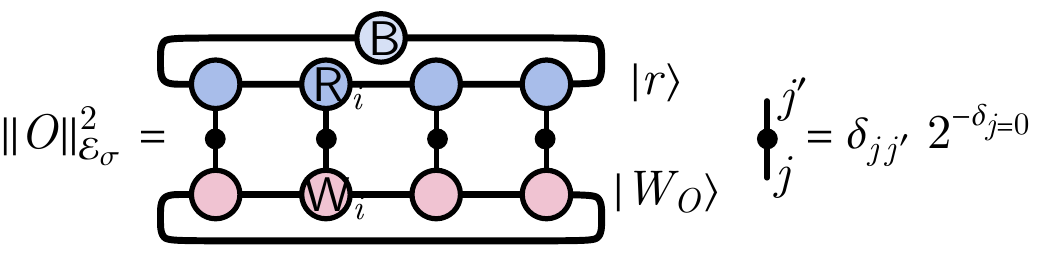}
    \caption{Tensor network representation of the locally-scrambled shadow norm $\|O\|_{\mathcal{E}_\sigma}^2$ defined in \eqnref{shadow-norm-general}.}
    \label{fig: shadow_norm_general_fig}
\end{figure}

Using the definition of the measurement channel $\mathcal{M}$ in \eqnref{eq: def M} and the reconstruction map $\mathcal{M}^{-1}$ in \eqnref{rcmap}, \eqnref{eq: def shadow norm} can be further reduced to
\begin{equation}\label{shadow-norm-general}
\begin{split}
\| O \|_{\mathcal{E}_\sigma}^2 &= 2^{-n}\Tr(O\mathcal{M}^{-1}(O))\\
&=\sum_A 2^{|A|-n}r_A (W_O)_A,
\end{split}
\end{equation}
where $r_A$ is the reconstruction coefficient given by the solution of \eqnref{consistency-eq}. $(W_O)_A$ is the entanglement feature of the operator $O$, which follows the definition of the entanglement features of a state ensemble in \eqnref{defEFstate}, with the density matrix $\rho$ simply replaced by the operator $O$:
\begin{equation}\label{defEFOperator}
(W_O)_A=\braket{A}{W_O}:=\Tr_A(\Tr_{\bar{A}}O)^2.
\end{equation}
It is also possible to represent $\ket{W_O}$ as an MPS, similar to the state entanglement feature in \eqnref{ef-mps-eq},
\begin{equation}\label{op-mps-eq}
    (W_O)_A=\Tr \prod_i (\delta_{i\notin A}\mathsf{W}_{O,i}^0+\delta_{i\in A}\mathsf{W}_{O,i}^1).
\end{equation}
This defines the MPS tensor $\mathsf{W}_{O,i}^j$ for any operator $O$. Hence, for a generic operator $O$, the shadow norm $\| O \|_{\mathcal{E}_\sigma}^2$ in \eqnref{shadow-norm-general} can be evaluated as a simple overlap given the MPS representations of two objects: the MPS matrices $\mathsf{R}^0,\mathsf{R}^1$ for the reconstruction coefficients $\ket{r}$ is given in \eqnref{r-mps-eq}; and the MPS matrices $\mathsf{W}_{O}^0, \mathsf{W}_{O}^1$ for  $\ket{W_O}$ as in \eqnref{op-mps-eq}. \figref{fig: shadow_norm_general_fig} illustrates the tensor network contraction, which can be computed efficiently with complexity linear in the system size.

The locally-scrambled shadow norm is upper bounded by the conventional shadow norm as $\| O \|_{\mathcal{E}_\sigma}^2\leq \|O\|^2_\text{shd}$. It plays a similar role in variance estimation, 
\begin{equation}\label{eq: variance_shadow_norm_relation}
    \mathrm{var}\,\langle O\rangle  \simeq \|O\|^2_{\mathcal{E}_\sigma}/M.
\end{equation}
The shadow norm $\|O\|^2_{\scE_\sigma}$ has two complementary physical meanings: (i) it characterizes the estimation variance $\mathrm{var}\,\langle O\rangle $ given a fixed number of samples $M$ (operators with larger shadow norms will have proportionally higher variances), (ii) it determines the sample complexity given the variance level (i.e.~$M\sim \|O\|^2_{\scE_\sigma}/\epsilon^2$ samples are needed in order to control the variance below a given threshold $\mathrm{var}\,\langle O\rangle \leq \epsilon^2$ set by a small $\epsilon$).

%

For Pauli estimation $O=P$, the operator entanglement feature $\ket{W_P}$ can be computed easily from its support. We define the support of a Pauli operator $P$ as the subset of sites on which it acts non-trivially, denoted as $\textsf{supp }P=\{ i | P_i\neq \id\}$. Define the \emph{operator weight} $k=|\textsf{supp }P|$ as the size of the support of $P$. We find that the operator entanglement feature of a Pauli string $P$ is simply given by
\begin{equation}\label{ef-Pauli-op-sets}
    (W_{P})_A = \delta_{\textsf{supp }P\subseteq A} 2^{2n-|A|}.
\end{equation}
The corresponding MPS matrices for Pauli observables $\mathsf{W}_{P,i}^0,\mathsf{W}_{P,i}^1$ take different values depending on if site $i$ is inside the support of $P$ or not. Given the general MPS form in \eqnref{op-mps-eq}, it is easy to check that
\begin{equation}\label{ef-Pauli-op}
    \mathsf{W}_{P,i}^0=4\delta_{i\notin \textsf{supp }P}, \qquad \mathsf{W}_{P,i}^1 = 2.
\end{equation}
Nonetheless, the matrices have bond dimension one because of the tensor product structure of $P$.

Plugging in \eqref{ef-Pauli-op-sets} into \eqref{shadow-norm-general}, we obtain the shadow norm of a Pauli string as a function of the reconstruction coefficients,
\begin{equation}\label{shadow-norm-pauli-r}
    \| P \|^2_{\mathcal{E}_\sigma} = 2^n \sum_{A|A\supseteq \textsf{supp }P} r_A.
\end{equation}
Using the solution of $\ket{r}$ in \eqnref{r-to-ef}, we have
\begin{equation}\label{shadow-norm-pauli}
     \| P \|_{\mathcal{E}_\sigma}^2 = \frac{(-3)^k}{\sum_{B |B\subseteq \textsf{supp }P} (-2)^{|B|}(W_{\mathcal{E}_\sigma})_B},
\end{equation}
where $k=|\textsf{supp }P|$. This result shows how the shadow norm of a Pauli operator $P$ depends on both its support and the entanglement feature of the snapshot state ensemble.

%

In the following, we will directly evaluate the locally-scrambled shadow norm $\|O\|^2_{\scE_\sigma}$ for several choices of the circuit depth $L$ following \eqnref{shadow-norm-general}. For notational convenience, we will treat $\ket{W_O}$ as translation-invariant unless evaluating the shadow norm for specific operator types that explicitly break translation symmetry, e.g. Pauli operators.

\subsubsection{$L=0$}
Here, the unitary ensemble is simply Pauli measurement. The vector $\ket{r}$ is a product states, and the reconstruction coefficients are given in \eqref{pauli-r} as $\mathsf{R}^0=-1,\mathsf{R}^1=3/2$. In terms of the matrices for $\ket{W_O}$, the expression for the shadow norm reduces the trace of a matrix power:
\begin{equation}\label{SN-L=0}
     \| O \|_{L=0}^2 = \Tr \left[\left( \frac{-\mathsf{W}_{O}^0 + 3\mathsf{W}_{O}^1}{2} \right)^n \right].
\end{equation}
We can see that in this limit, the shadow norm is closely related to the locality of the operator, since if $O$ is the identity on some site $i$, then the corresponding contribution to the shadow norm is a factor of $1$, i.e.~it doesn't affect the shadow norm. 

Consider the case $O=P$. In the translation-symmetry breaking version of \eqnref{SN-L=0}, the power of $n$ breaks into individual factors for each site, with each term in the product depending on the operator entanglement feature matrix at that site:
\begin{equation}
     \| P \|_{L=0}^2 =\prod_{i=1}^n \left( \frac{-\mathsf{W}_{P,i}^0 + 3\mathsf{W}_{P,i}^1}{2} \right).
\end{equation}
Using \eqnref{ef-Pauli-op}, the product contributes a factor of 1 where $P_i$ is identity and a factor of 3 where $P_i$ is not. Hence we see that the locally-scrambled shadow norm for $P$ grows exponentially in the weight $k=|\textsf{supp }P|$,
\begin{equation}
     \| P \|_{L=0}^2 = 3^{k}.
\end{equation}
This result agrees with the conventionally defined shadow norm. It can also be seen to agree with \eqref{shadow-norm-pauli} since the entanglement feature state of the unitary ensemble $\ket{W_{\mathcal{E}_\sigma}}$ is a trivial product state in the $L=0$ case. 

Furthermore, although the locally scrambled shadow norm is less than the conventionally defined shadow norm, for locally-scrambled unitary ensembles, it provides accurate bounds on the single-shot variance of Pauli observables \cite{Hu2107.04817}. Support for this claim is given in \secref{numerics}.

\subsubsection{$L=1$}\label{sec: L=1}

Another important case we may check is $L=1$. In this case, the ensemble is a product of two-local, two-design unitary gates. Each gate effectively scrambles site $2i$ with its neighbor at $2i+1$, and generates no entanglement outside this unit cell. Within each unit cell, the unitary is Clifford random. Thus the MPS for reconstruction coefficients in each block for the $L=1$ case is given by \eqref{clifford-r}. Therefore, 
\begin{equation}\label{SN-L=1}
    \| O \|_{L=1}^2 = \Tr \left[ \left( \frac{-(\mathsf{W}_{O}^0)^2 + 5(\mathsf{W}_{O}^1)^2}{4} \right)^{n/2} \right].
\end{equation}
Consider next the case $O=P$, and suppose $n$ is even for simplicity. Then the translation-symmetry breaking version of \eqref{SN-L=1} for a Pauli operator is
\begin{equation}
    \| P \|_{L=1}^2 = \prod_{i=1}^{n/2} \left( \frac{-\mathsf{W}_{P,2i-1}^0 \mathsf{W}_{P,2i}^0 + 5\mathsf{W}_{P,2i-1}^1 \mathsf{W}_{P,2i}^1}{4} \right).
\end{equation}
There are two types of factors depending on the support of the operator $P$ in that unit cell. If the Pauli operator $P$ is trivial in the unit cell, then the factor is one. Otherwise, if the Pauli operator $P$ has any support in the unit cell, then the factor is five. Hence, for a contiguous Pauli operator of weight $k$, there is a (staggered) exponential growth in the shadow norm due to the unit cell structure of the brick wall circuit i.e.
\begin{equation}
     \| P \|_{L=1}^2 = 5^{\lceil k/2 \rceil}.
\end{equation}
Thus, as the operator weight $k$ grows, the shadow norm, and hence the number of samples required for reliable prediction, is already exponentially smaller for $L=1$ than $L=0$, since $\frac{\| P \|_{L=1}^2}{\| P \|_{L=0}^2}\sim (\frac{\sqrt{5}}{3})^{k}$. We find that this pattern, that longer operators can be better estimated by deeper circuits, continues for $L>1$ in \secref{numerics}. For a given operator weight $k$, there is an ideal short circuit depth $L^*\geq 0$ such that the shadow norm and, therefore, the singles-shot variance are minimized. This shows that classical shadow tomography beyond Pauli measurements has practical applications since we can use circuits of varying depths to minimize the number of samples one has to take. 

When $L\geq 2$, the shadow norm is more difficult to evaluate analytically. This is because both the entanglement features of the state ensemble and the reconstruction coefficients are more complicated. While the cases $L=0,1,\infty$ can be helpful, for $1 < L < \infty$, we resort to numerical evaluation of the shadow norm. See \secref{numerics}. We find that the shadow norm is a reliable indicator of the estimation variance in the case where $O$ is a tensor product of local operators.

\subsubsection{$L=\infty$}

When $L\rightarrow \infty$, there are only two non-zero reconstruction coefficients $r_{\emptyset}=-1$ and $r_{\Omega}=1+2^{-n}$, which correspond to the empty set and the total system, respectively. Then the shadow norm $ \| O \|_{L=\infty}^2 $ for a traceless Hermitian operator reduces to 
\begin{equation}
   \| O \|_{L=\infty}^2 = (1+2^{-n})\Tr(O^2). 
\end{equation}
This is reminiscent of the conventional shadow norm introduced in \cite{Huang2002.08953}, as both are proportional to $\Tr (O^2)$, indicating that the locally-scrambled shadow norm in the deep circuit limit also depends strongly on the rank of the operator. However, the locally-scrambled shadow norm has a smaller proportionality factor since the conventional shadow norm is defined by taking a maximum over all state $\rho$. We also note that afterwards in \cite{arienzo2022closedform}, some closed form analytic expressions were developed for the reconstruction map in brickwall circuits that are consistent with these results.

\section{Numerical Demonstrations}\label{numerics}

In this section, we demonstrate the prediction algorithms for various system sizes based on numerical simulation\footnote{Most of the numerical codes used in this paper can be found at this \href{https://github.com/Ahmed-Akhtar/ShallowCircuitClassicalShadows}{github} repository.}. We evolve the system using random two-local Clifford gates arranged in a brick wall pattern with periodic boundary conditions. The depth $L$ of the circuit refers to the number of layers of Clifford gates. For example, for $L=2$, we apply one even layer and then one odd layer of independently random two-qubit local Clifford gates. After the Clifford circuit evolution, we then measure the qubits and store the resulting snapshot state $\hat{\sigma}=U^\dagger \ket{b}\bra{b} U$ for later use in predicting various quantities. For demonstration purposes, we collect the classical snapshots by numerically simulating the randomized measurement. Suppose the underlying original state $\rho$ is a stabilizer state; the simulation can be efficiently carried out on a classical computer using the stabilizer table algorithm, thanks to the Gottesmann-Knill theorem. 

In our numerical demonstrations, we consider two example initial states: the cluster state $\rho_\text{ZXZ}$ and the Greenberger-Horne-Zeilinger (GHZ) state $\rho_\text{GHZ}$. They can be specified by their stabilizer groups, 
\begin{align}
    \mathcal{G}_\text{ZXZ} &= \langle Z_1 X_2 Z_3, \cdots, Z_n X_1 Z_2 \rangle, \\
    \mathcal{G}_\text{GHZ} &= \langle Z_1 Z_2, \cdots, Z_{n-1}Z_n, \textstyle\prod_{i=1}^{n}X_i\rangle,
\end{align}
Each stabilizer state $\rho={|\mathcal{G}|}^{-1}\sum_{g\in\mathcal{G}} g$ is defined as the invariant state of the corresponding stabilizer group $\mathcal{G}$.

In the following figures, the initial state will be indicated using squares (for $\rho_\text{ZXZ}$) and triangles (for $\rho_\text{GHZ}$), respectively. The cluster state $\rho_\text{ZXZ}$ is the ground state of a gapped, local, stabilizer Hamiltonian, and therefore carries strictly area-law entanglement. The GHZ state $\rho_\text{GHZ}$ is the symmetric ground state of the Ising model, which contains long-range entanglement and is more challenging for prediction based on the randomized measurement on shallow local circuits. For example, the extensive generators (i.e.~$\prod_i X_i$ in the $\mathcal{G}_\text{GHZ}$ stabilizer group) will be hard to probe by local circuits of a finite depth $L$, because the generators of the snapshot state are at most $\sim 2L$ in length.

We then use the reconstruction coefficients, together with the algorithms outlined above, to perform prediction using the snapshot states. The reconstruction coefficients are calculated using PyTorch, specifically the AdamW optimizer, by minimizing the loss function $\mathcal{L}$ in \eqnref{eq: loss}. The minimization procedure is continued until $\mathcal{L}\leq 10^{-3}$.

We would like to briefly comment on the flexibility of this approach with respect to different circuit evolutions. As explained in \figref{fig:CST}, the input for determining the reconstruction coefficients is the entanglement features of the underlying unitary ensemble. One might be wondering if the brick-wall Clifford unitary circuit is the only specialized case which admits efficient entanglement feature dynamics. For example, one may be interested doing tomography using other locally-scrambled unitary ensembles, such as random Hamiltonian-generated evolution, which was proposed in \cite{Hu2107.04817}. It was found in prior work that locally-scrambled dynamics generated by continuously scrambling gates is also well described by a simple $D=2$ MPS ansatz \cite{Akhtar2006.08797}. Similar efficiency was found in \cite{Fan2002.12385}. It is believed that this locally-scrambled purity dynamics is generically efficiently representable as MPS away from finely tuned points like the measurement-induced entanglement transition. Therefore, the shallow-circuit tomography scheme presented in this paper can be adapted to other circuits which are also locally-scrambled at the two-design level. We will leave the analysis of the tomography scheme in these more general contexts to future work.

Lastly, a similar approach using MPS to do efficient shallow circuit tomography was developed concurrently to ours in \cite{Bertoni2209.12924}. In this paper, the Pauli weight instead of the entanglement feature is computed. Since these quantities are related by local basis transformation in the entanglement-feature Hilbert space, we suspect that the efficiency of their approach will be comparable to ours.

\subsection{Pauli Estimation}

To demonstrate the efficiency of our approach in estimating Pauli observables, we will take $Z^{\otimes k}:=\bigotimes_{i=1}^k Z_i$ as the target observable without loss of generality. The operator weight (operator size) $k$ can be adjusted. Since the randomized measurement schemes are invariant under local Clifford transformations, any Pauli string $P$ of the same weight $k$ will exhibit the same sample efficiency (as the shadow norm $\| P\|_{\scE_\sigma}^2$ only depends on the support of $P$). Nevertheless, the expectation value $\langle P\rangle$ will still depend on the choice of the observable $P$ and the underlying state. On the GHZ state, the expectation value of $Z^{\otimes k}$ is 1 or 0 depending on if $k$ is even or odd, respectively. On the cluster state, it is 0 for all $k\geq 1$.  

\begin{figure}[htbp]
    \centering
    \includegraphics[width=0.72\columnwidth]{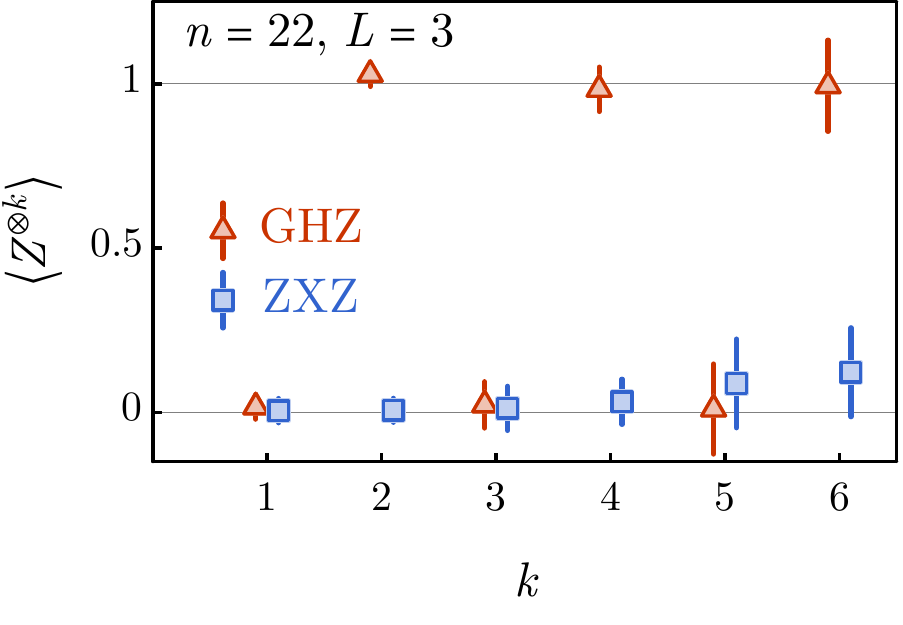}
    \caption{The estimated Pauli observable $Z^{\otimes k}:=\bigotimes_{i=1}^k Z_i$ of different weight $k$ with the underlying state $\rho$ being the cluster (squares) and GHZ (triangles) states. Each point is based on 50000 measurement samples collected from a brick-wall circuit of depth $L=3$ measurement protocol. The system size is $n=22$ qubits. The error bar indicates two standard deviations of the estimated value.
    }
    \label{fig:pauli_prediction}
\end{figure}

In \figref{fig:pauli_prediction}, we demonstrate that our method can reliably predict the expectation value of the Pauli string observable $Z^{\otimes k}$ on a system of $n=22$ qubits via randomized measurements on the random Clifford circuit of $L=3$ layers. Our numerical result is obtained by first simulating the randomized measurement on a classical computer and then processing the data using the algorithm described in \secref{sec: pauli}. Our classical post-processing approach provides unbiased estimations of $Z^{\otimes k}$ on both the cluster state (blue squares) and the GHZ state (red triangles) for various operator weights $k$, although the estimation variance is growing with $k$ (which will be analyzed soon). Here we only demonstrate the case of circuit depth $L=3$, but our method is applicable for other circuit depths, and the quality of estimation remains similar.

\begin{figure}[htbp]
    \centering
    \includegraphics[width=0.75\columnwidth]{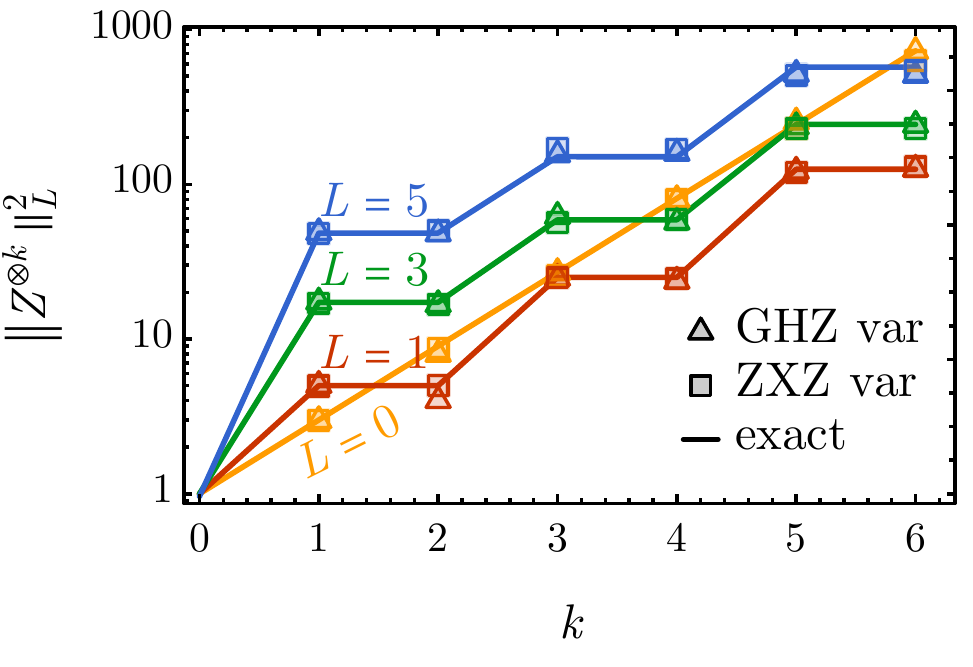}
    \caption{The locally-scrambled shadow norms for circuit depth $L=0,1,3,5$ (yellow, red, green, and blue lines, respectively), with $n=22$ qubits. We also plot the corresponding variances over all samples for the cluster (squares) and GHZ (triangles) states. These agree with the shadow norms. The staggered behavior of the shadow norm for $L>0$ is because these circuits break translation invariance.}
    \label{fig:pauli_variance}
\end{figure}

The variance of the estimation is associated with the sample complexity. A larger variance (a larger shadow norm) means more samples are needed to reduce the variance to the desired accuracy level. The traditional classical shadow tomography with Pauli measurements ($L=0$) quickly requires an extensive number of samples to estimate a weight-$k$ Pauli string observable (as the shadow norm $\| P \|_{L=0}^2 = 3^{k}$ grows exponentially with $k$). However, with shallow circuits $L\geq 1$, we can easily achieve the same accuracy with far fewer samples. We have analytically proved this statement in \secref{sec: L=1} for $L=1$ by calculating the locally-scrambled shadow norm. For large $L$, it is difficult to obtain a closed-form analytic expression for the shadow norm, but we can numerically investigate the shadow norm as a function of $L$ according to \eqnref{shadow-norm-general} using tensor network techniques. Our results are presented in \figref{fig:pauli_variance}.

We first verify that the locally-scrambled shadow norm $\|Z^{\otimes k}\|_L^2$ computed from \eqnref{shadow-norm-general} correctly captures the variance $\mathop{\mathrm{var}}Z^{\otimes k}$ of the estimated expectation value observed in our numerical simulation. Based on \eqnref{eq: variance_shadow_norm_relation}, we expect the variance and the shadow norm to match each other, as $\mathop{\mathrm{var}} \, \langle Z^{\otimes k} \rangle = \|Z^{\otimes k}\|_L^2  / M$, where $M$ is the number of samples. This is indeed the case as shown in \figref{fig:pauli_variance}. The numerically simulated variance and the computed shadow norm are in good agreement as we vary $L$ and $k$, even as $k$ approaches a quarter of the system size. This confirms the correctness of our approach to computing the shadow norm. The shadow norm is useful as it can give us an idea of the number of samples needed to control the estimation variance to the desired level, even before we start performing the classical shadow tomography experiment. 

As shown in \figref{fig:pauli_variance}, the shadow norm $\|Z^{\otimes k}\|_L^2$ always grows with the operator weight $k$, meaning that we always need to perform more measurements to estimate longer Pauli strings accurately. However, the growth rates of shadow norms are different for different circuit depths $L$. For example, the $L=1$ shadow norm increases with $k$ slower than that of $L=0$, meaning that the $L=1$ circuit will be more advantageous for probing Pauli string operators of sufficiently large weight $k$. 

\begin{figure}[htbp]
    \centering
    \includegraphics[width=0.75\columnwidth]{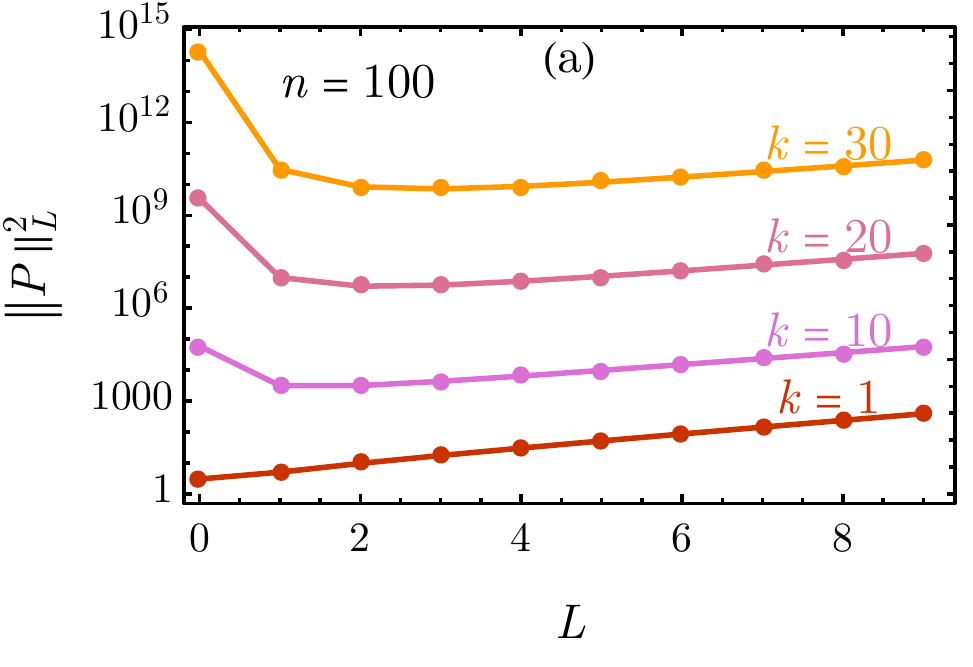} \\
    \includegraphics[width=0.75\columnwidth]{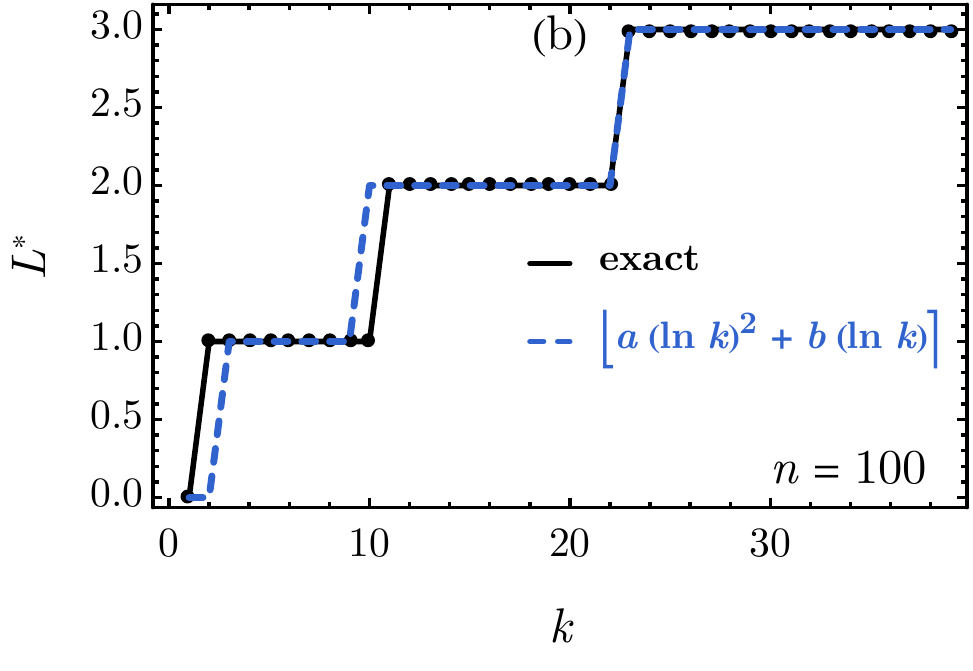}
    \caption{(a) The locally-scrambled shadow norm for a general Pauli operator $P$ of size $k$ v.s. the circuit depth $L$ of the randomized measurement scheme. (b) The optimal value of circuit depth $L^{*}$ to predict a contiguous Pauli operator of weight $k$. 
We use a system size of $n=100$ qubits here. The results do not change as we further increase $n$. 
}
    \label{fig:pauli_shadow_norm}
\end{figure}

To study this trend more quantitatively, we use \eqref{shadow-norm-pauli} to compute the shadow norm of Pauli operators for various weights $k$ under various circuit depths $L$. The result is presented in \figref{fig:pauli_shadow_norm}(a). Given the Pauli operator size $k$, it is clear that there is an optimal circuit depth $L^*$ that minimizes the shadow norm. We conjecture that $L^*$ can be fitted by an empirical formula
\eq{\label{eq: L* fitting}
L^*\sim a (\ln k)^2 + b (\ln k).}
The formula fits the numerically found $L^*$ nicely with parameters $a=0.14,b=0.35$, as shown in \figref{fig:pauli_shadow_norm}(b), where the rounded value of the fitting function is shown since $L^*$ must always be an integer. This confirms the sub-polynomial growth in the optimal circuit depth $L^*$ as proposed in \eqnref{eq: L* fitting}.  The result indicates that a relatively shallow circuit can efficiently optimize the sample complexity for classical shadow tomography of large Pauli strings, which demonstrates the advantage of our shallow-circuit measurement approach. The logarithmic scaling of the optimal circuit depth was confirmed afterwards in \cite{ippoliti2023operator}. In this work, the authors were able to derive a bound on the shadow norm by studying the evolution of the weight distribution of contiguous operators under the random brick-wall circuit.


We would like to point out that for estimating the Pauli string with a fixed operator weight $k$, the shadow norm remains constant with the system size $n$, meaning that classical shadow tomography is scalable for this task. Our approach of using shallow-circuit measurements enables further improvement in the sample complexity compared to the Pauli measurement without losing the scalability and feasibility on near-term quantum devices.

\subsection{Fidelity Prediction}

Fidelity estimation is another important task in quantum information. We can also compare the fidelity of $\rho_\text{GHZ}$ and $\rho_\text{ZXZ}$ against various states using classical shadow tomography on shallow circuits. 
A high fidelity $F=1$ ensures that the state produced in the lab is close to the desired state, whereas a low fidelity means that the state produced in the lab has very little overlap with the desired state. To demonstrate our approach, we simulate the shallow-circuit randomized measurement on a classical computer and then use the algorithm outlined in \secref{sec: fidelity} to process the measurement data collected from the simulation. We estimate the fidelity of the classical shadow reconstructed state with the underlying original state and show that our approach successfully reconstructs the original quantum state from measurement outcomes with high fidelity. Although the shallow-circuit classical shadow tomography still has exponential sample complexity for the fidelity estimation task, increasing the circuit depth a bit can significantly reduce the base of this exponential complexity, which makes our approach useful for quantum state tomography on near-term quantum devices.

\begin{figure}[htbp]
\begin{center}
\includegraphics[width=0.73\columnwidth]{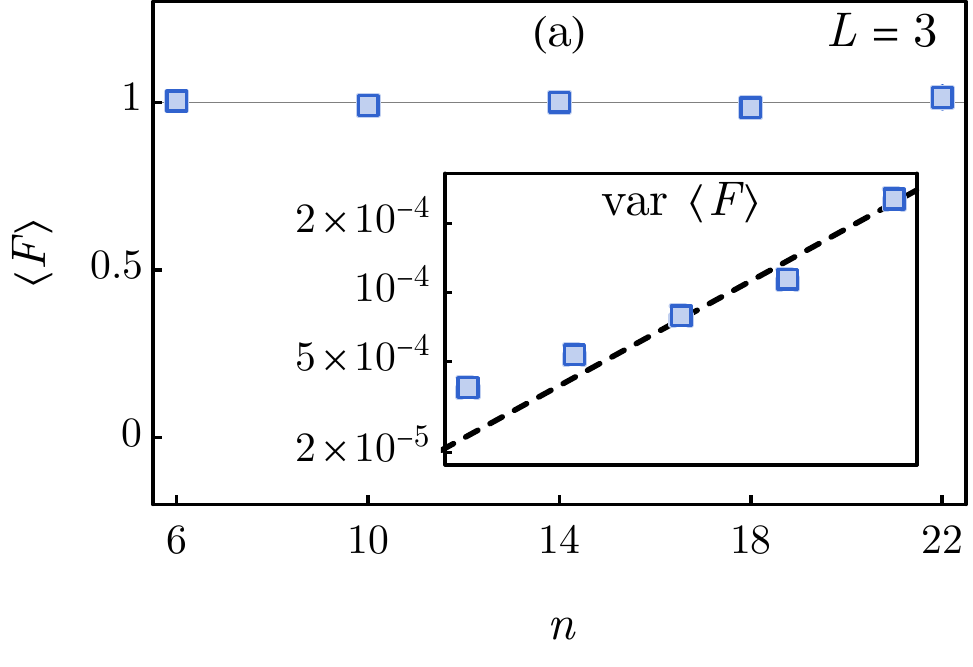} \quad
\includegraphics[width=0.73\columnwidth]{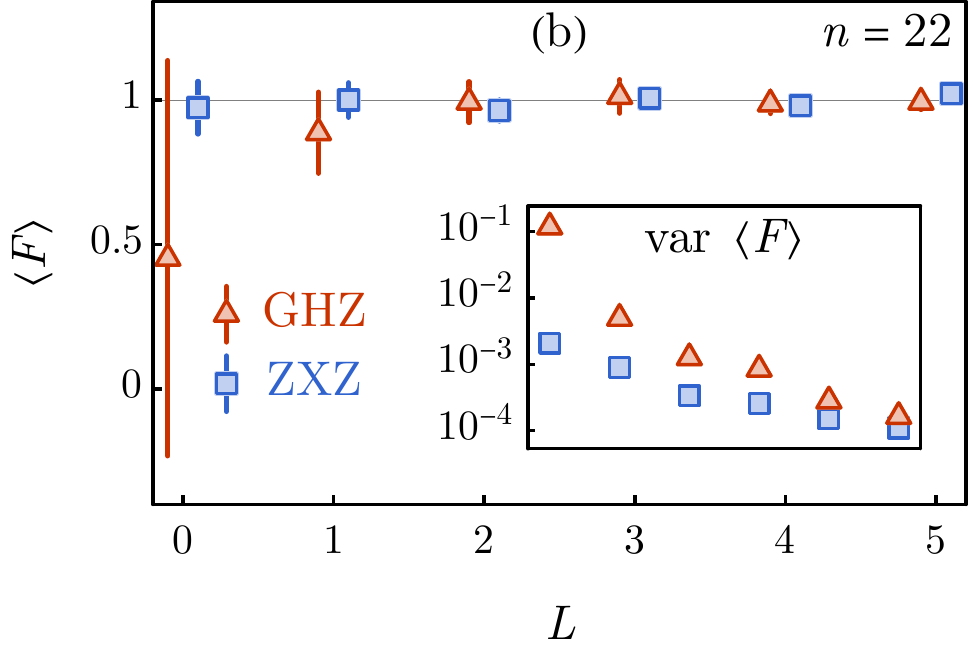}
\caption{(a) The fidelity and its variance as a function of system size, for systems of size $n=6,10,14,18,22$ for a fixed circuit depth of size $L=3$. We use the cluster state as the reference state. Each point is the average of 50000 measurement samples. The mean fidelity is tightly centered around one, which is expected. The growth of the sample variance is indicated in the inset. 
(b) The fidelity and its variance versus circuit depth $L$ for a system of $n=22$ qubits. The two states are the cluster state (squares) and the GHZ state (triangles). We use a median-of-means average to estimate the fidelity, using 50000 total samples broken up into 12 groups. The error bar corresponds to two standard deviations. The inset shows that the variance decays with $L$ rapidly.}
\label{fig: fid_pred}
\end{center}
\end{figure}

First, we compare the fidelity of the cluster state with itself as a function of the qubit number $n$ for $n=6,10,14,18,22$, using a circuit depth of $L=3$, using 50000 samples. The result is shown in \figref{fig: fid_pred}(a). We find the reconstructed states have close-to-one fidelities with negligible error bars up to 22 qubits. We also observe a (rather slow) exponential growth in the variance of the fidelity, in agreement with \refcite{Hu2107.04817}. In that paper, a scaling form is proposed with $\text{var }F \propto \exp(c n / (L+1)^\alpha)$. 
Fixing $\alpha=0.72$, per \refcite{Hu2107.04817}, we can fit the parameter $c$ based on our data.
We find that the fitted value $c=0.4$ agrees with the value $c=0.47\pm 0.08$ previously observed in \refcite{Hu2107.04817}. This empirical relation suggests that although the variance of fidelity estimation (and hence the sample complexity) grows exponentially with the number of qubits $n$, increasing the circuit depth $L$ in the shallow circuit regime can significantly reduce the variance (by making the exponential growth very slow). 

Next, we study the efficiency of the protocol as a function of the circuit depth $L$ for a system of size $n=22$, as shown in \figref{fig: fid_pred}(b). We plot the fidelity of both the cluster state and the GHZ states with themselves and find that both are centered around one, as expected. For both states, we observe a rapid decline in the variance of the sample fidelity as we increase circuit depth. However, we notice that the GHZ state has a much higher sample variance. This is likely due to the fact that the GHZ state has long-range correlations, unlike the cluster state which only has finite-range correlations. Thus quasi-local measurements will not be very efficient in probing the GHZ state. For example, at $L=0$ (Pauli measurements), the reconstructed state has very low fidelity $F<0.5$ and very large sample variance $\text{var} F\sim 1$ for the GHZ state. We see a sharp (more than an order of magnitude) decline in the sample variance for $L=1$, and a continued exponential decline as we increase $L$. By $L=3$, the fidelity approaches one for both states within a variance less than $10^{-3}$. This is because a deeper scrambling circuit will enable the final computational basis measurements to probe more complex (higher weight) operators in the original basis, which benefits the fidelity estimation. This demonstrates the great advantage of entanglement-assisted shallow-circuit measurements. The rapid decline of the variance in fidelity versus circuit depth $L$ makes it promising to use classical shadow tomography for fidelity estimation with shallow circuits achievable on near-term quantum devices.

\section{Conclusion and Open Questions}\label{conclusion}

In this paper, we demonstrate a novel approach to predicting expectation values of observables using shallow-depth circuit classical shadow tomography. Our work extends the well-known Pauli measurement and Clifford measurement protocols to the case of finite-depth brick wall circuits, which are available on near-term devices on many qubits. This allows us to use shallow circuit measurement to ``approximate'' the Clifford measurement limit to do fidelity estimation efficiently. It also allows us to use the scrambling properties of circuits with $L>0$ to more efficiently probe large-weight operators, e.g.~long Pauli strings. The approach that we take is based on our entanglement features formalism, which allows us to simply represent the purity data as MPS and consequently determine the reconstruction coefficients. Given the reconstruction coefficients and that they can be represented with a low-bond dimension MPS, we propose a scalable approach to predict many properties of quantum states.

There are still several questions that remain and that we wish to explore in future works. Firstly, is there an exact and efficient representation of the reconstruction coefficients as MPS? In our paper, we determine the reconstruction coefficients by learning their MPS tensors via gradient descent on the consistency equation \eqnref{consistency-eq}. This procedure only needs to be done once, but it is bound to produce some error that can propagate through the prediction algorithm. In \refcite{Bu2022}, the exact, individual reconstruction coefficients were determined. In practice, we want to be able to organize all $2^n$ coefficients efficiently --- is there some way to represent the exact individual coefficients scalably? Such an approach would also give us insight into how the bond dimension or complexity of the reconstruction map evolves with circuit depth. We know that it is small in the shallow circuit limit and also small at $L=\infty$. Is it possible that the bond dimension remains small and finite for all values of the circuit depth $L$? Furthermore, is there some way to generalize the results in this paper to higher dimensions? For example, do the entanglement feature states and reconstruction coefficients now a low-dimensional PEPS structure?

Moreover, the prediction procedures described in \secref{algorithms} assume that the operator whose expectation value we want to predict has an efficient Pauli basis expansion. This means that the operator with a large amount of operator entanglement will be difficult to predict. Furthermore, this procedure is limited by the growth in operator entanglement of the snapshot states over time. The snapshot states' bond dimension grows exponentially in $L$, which can be seen from the fact that the typical size of the generators of the snapshot states grows linearly in $L$. This means this procedure will be costly for deep circuits. What approach can we take in this case?

In addition, we only considered measuring quantities that depend on the first moment of $\rho$ i.e. $\Tr(O\rho)$. We could also consider quantities that depend on higher moments of $\rho$, e.g. $\Tr(O \rho\otimes \rho)$, where $O$ is an operator on the doubled Hilbert space. How can we generalize the approach developed here to expectation values based on higher moments of $\rho$?

In this paper, we only considered a particular class of circuit ensembles formed by varying the circuit depth $L$. However, the procedure we outlined is independent of the particular circuit geometry. How might other geometries be useful in other contexts? For example, instead of considering depth $L$ circuits, we could also consider circuits composed of adjacent $k$-qubit unitaries. Such circuits should also be feasible on near-term devices, provided that $k$ is small. More ambitiously, one could also consider circuits with holographic geometries. Such circuit geometries may be optimal for prediction on gapless or scale-invariant states. We leave these cases to future research.

There are still many conceptual questions that have yet to be answered. For example, what is the physical meaning of the locally-scrambled shadow norm? The norm is closely related to the scrambling properties of the unitary ensemble, which begs the question of how it can be related to other information-theoretic and thermodynamic quantities. 

Our setup assumes a qubit system, where the local degrees of freedom are finite and commuting. How do we apply our approach in \secref{algorithms} to the fermionic context or other experimental contexts which may not be easily described in terms of qubits? 

One may also consider different types of locally-scrambled or approximately locally-scrambled ensembles than the Clifford ensemble. For example, one may consider instead quantum Brownian motion ensembles, where the evolution is generated by a random Hamiltonian coupling nearby sites. The entanglement features of such ensembles have been extensively studied in prior work \cite{You1803.10425, Kuo1910.11351, Akhtar2006.08797}. What are the reconstruction coefficients for such ensembles? What are their prediction properties? Classical shadow tomography is a rapidly developing field with many interesting applications and open questions. We hope that our paper provides some insights that can motivate further interesting developments.

\begin{acknowledgements}
We acknowledge the discussions with Christian Bertoni, Jonas Haferkamp, Marcel Hinsche, Marios Ioannou, Jens Eisert, Hakop Pashayan, who were working on similar topics concurrently \cite{Bertoni2209.12924}. We thank Soonwon Choi for the helpful discussion and previous collaboration. We also thank Xun Gao for the helpful discussion on tensor networks. The authors are supported by the UCSD Hellman Fellowship. This research was done using services provided by the OSG Consortium \cite{osg07,osg09}, which is supported by the National Science Foundation awards 2030508 and 1836650.

\end{acknowledgements}

\bibliographystyle{plainnat}
\bibliography{ref}

\onecolumngrid

\newpage
\appendix

\section{MPS-representation of a finite-depth stabilizer states}\label{mps-stabs}

Here, we outline an algorithm for constructing the MPS representation of a stabilizer state, given the generators $g_i$ of the corresponding stabilizer group $G=\langle g_1 \cdots  g_n \rangle$. In the classical shadow protocol, we start with a trivial product state $\ket{b}=\otimes_{i=1}^{n}\ket{b_i}$ which represents a measurement outcome in the computational basis,  and evolve by a two-local Clifford circuit. The initial stabilizers therefore can transform into more complicated Pauli group elements.

\begin{equation}
    (-1)^{b_i} Z_i \rightarrow g_i  
\end{equation}

For a Clifford circuit of depth $L>0$, the size of $g_i$ is at most $2L$. The state $\rho$ is then a sum over stabilizer group elements. Similarly, we can also represent the state as an MPS in the Pauli-basis, described by coefficients $c_{\vec{s}}$ where $\vec{s}\in \{ 0\cdots  3\}^n$ denotes a Pauli string.

\begin{equation}
    \rho = \frac{1}{2^n} \sum_{g\in G} g =  \frac{1}{2^n} \sum_{\vec{b} \in \Omega^n } \prod_{i=1}^n g_i^{b_i} = \frac{1}{2^n}\sum_{\vec{s}} c_{\vec{s}}P^{\vec{s}}
\end{equation}

For example, the GHZ state on two sites has two generators, $g_1=ZZ,g_2=XX$, and four nonzero components in the Pauli basis, corresponding to $II,XX,YY,ZZ$. Therefore, there are four nonzero components for $c$, namely $c_{00}=c_{11}=c_{33}=-c_{22}=1/4$, and all other components are zero. 

We describe a procedure to determine $c_{\vec{s}}$ as an MPS, given a set of generators $g_1\cdots g_n$ and corresponding phases $(-1)^{s_1}\cdots (-1)^{s_n}$. The idea is that the local MPS tensor at site $x$ should transmit through its virtual indices which of the stabilizers that have support on $x$ are included in the product (i.e. have $b_i=1$). First, we need to determine a contiguous set of sites that contain the support of each generator. Let us call this the \textit{extent} of an operator. Given the extent of each generator, let $P_i=\mathbb{I}+(-1)^{s_i}g_i$ be an projection operator (or state in the Pauli basis) defined on the extent. Schematically, the local MPS tensor $[P_i^j]$ at site $j$ is 

\begin{equation}
    [P_i^j] = \begin{cases} 
        (\mathbb{I},(-1)^{s_i}g_i^j) & j=1 \\
        \begin{pmatrix} \mathbb{I} & 0 \\ 0 & g_i^j \end{pmatrix} & 1<j<k \\
        (\mathbb{I},g_i^j)^T & j=k
    \end{cases} 
\end{equation}

where $k$ is the size of the extent, and $g_i^j$ is the Pauli operator at site $j$ in $g_i$. Given the MPS $P_i$, we only need one more ingredient to determine the MPS representation of the stabilizer state: Pauli algebra fusion tensor $V^c_{a,b}$. This tensor implements the product of Pauli operators, and evaluates to the correct phase associated to the product e.g.

\begin{equation}
    V^c_{a,0} = \delta_{a,c} , V^{3}_{1,2}=i=-V^{3}_{2,1}, \textsf{etc.}
\end{equation}

Then, to determine the MPS tensor for the stabilizer state at site $j$, first collect all the generators whose extent is on $j$. Then, combine their MPS along the physical dimension using the fusion tensor. If there is more than two generators with support at $j$, then we can use multiple fusion tensors.

\end{document}